\documentclass[aps,prd,twocolumn,groupedaddress,amsmath,amssymb,floatfix]{revtex4-2}
\usepackage{lineno,graphicx,color,hyperref,bm, orcidlink,placeins}

\begin{document}

\title{Nuclear deexcitation simulator for neutrino interactions and nucleon decays \\ of $^{12}$C and $^{16}$O based on TALYS}

\newcommand{\kamioka}{\affiliation{Kamioka Observatory, Institute for Cosmic Ray Research, University of Tokyo, Kamioka, Gifu 506-1205, Japan}}
\author{Seisho Abe\orcidlink{0000-0002-2110-5130}} \kamioka
\date{\today}

\begin{abstract}
Nuclear deexcitation associated with neutrino interactions and nucleon decays has a significant impact on a recent key observable for neutrino detectors: neutron multiplicity.
Although several deexcitation simulation studies have been conducted, most are closed-source or have limitations in application.
Therefore, we need an open-source deexcitation simulator to be used as a standard deexcitation module in existing neutrino Monte Carlo event generators.
To predict the neutron multiplicity comprehensively, the author developed a dedicated nuclear deexcitation simulator based on TALYS named NucDeEx.
NucDeEx can be used for neutrino interactions and nucleon decays for $^{12}$C and $^{16}$O.
The author provides the NucDeEx codes and sample codes for applying the NucDeex to widely used neutrino Monte Carlo event generators, NEUT, NuWro, and GENIE.
\end{abstract}
\maketitle

\section{Introduction} \label{sec:introduction}
Neutrino-nucleus interactions are frequently described by the impulse approximation, which divides the process into three parts:
Initial nucleon state, primary interaction, and final state interactions (FSI).
The initial nucleon state is usually parametrized by the Fermi gas model or spectral functions.
The primary interaction is the process via the weak interaction of a neutrino with a single nucleon target, treating the other nucleons as mere spectators.
The FSI is a cascade rescattering of particles in the nucleus.
The description above is adopted in most of the existing neutrino Monte Carlo event generators.
Considering actual interaction processes, a hole or holes are created after the FSI.
Then, the residual nucleus frequently remains in the excited state and undergoes deexcitation, emitting particles such as protons, neutrons, $\alpha$ particles, and gamma rays.
However, with few exceptions~\footnote{NEUT considers simplified deexcitation process based on Ref.~\cite{kobayashi2006deexcitation} only for $^{16}$O target.},
this process was not simulated in the major generators until the recent studies described later.
Despite this fact, this description has been used for a long time in various neutrino experiments with much success because the particles emitted by nuclear deexcitation have low energies of a few MeV and, except for gamma rays, are undetectable with most neutrino detectors.
\par
\begin{figure}[htbp]\centering
\includegraphics[width=0.85\columnwidth]{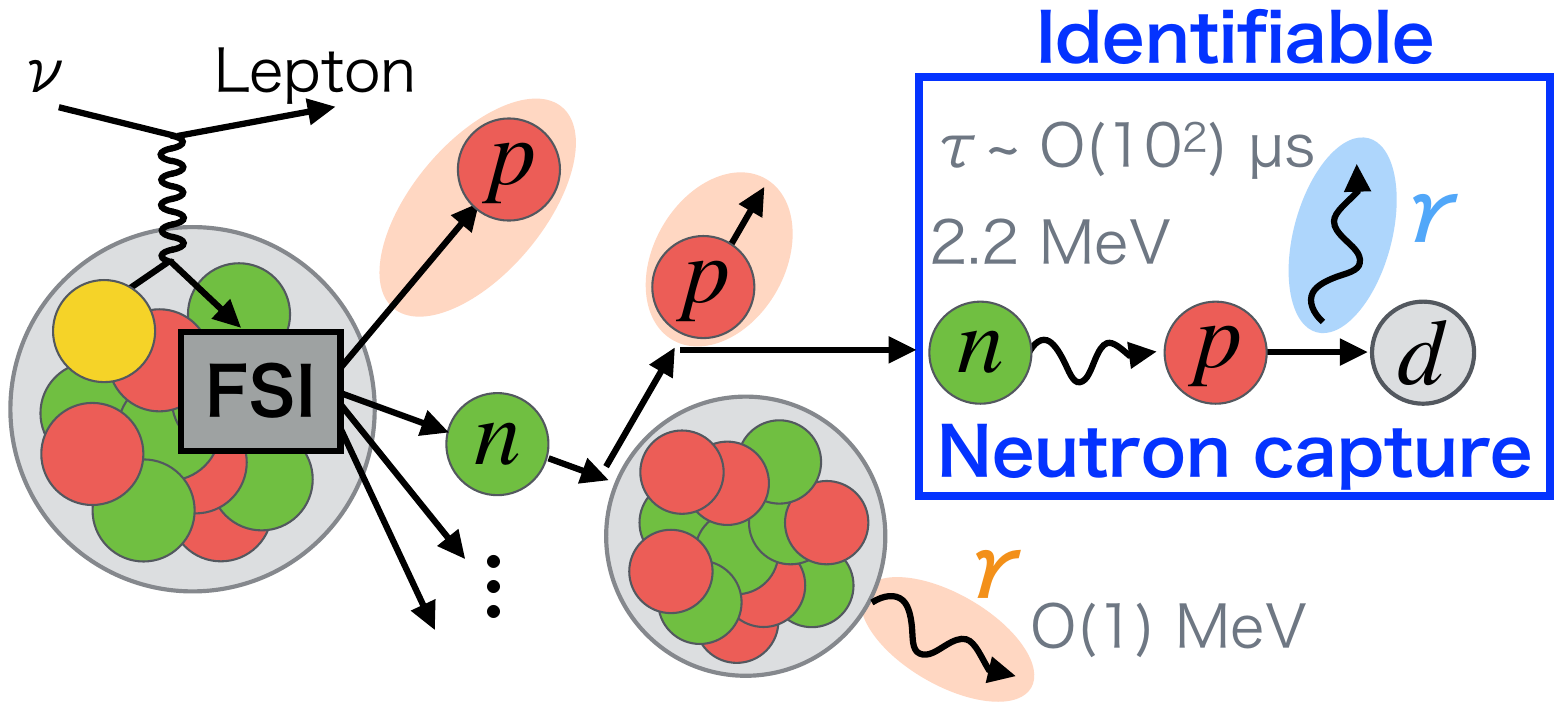}
\caption{Schematic view of detection of neutrino-nucleus interactions.
         Neutrons recoil protons, leaving short trajectories of a few centimeters.
         Recoiled nuclei sometimes emit deexcited gamma rays with $\mathcal{O}(1)$\,MeV quickly.
         After thermalization, neutrons are captured mostly by protons (gadolinium) emitting 2.2\,MeV ($\sim8$\,MeV) gamma rays with a particular lifetime of about several hundred microseconds.
         Neutrons are detectable and identifiable by detecting the gamma ray without setting energy thresholds for neutrons.
         }
\label{fig:detection}
\end{figure}

Neutrino detectors can detect neutrons in three ways: Detection of recoil protons, deexcited gamma rays from recoiled nuclei, and gamma rays emitted by neutron capture, as shown in Fig.~\ref{fig:detection}.
Since the protons have a high Cherenkov threshold and short trajectories of a few centimeters, Cherenkov and tracker detectors are forced to impose high detection thresholds.
Because of their low energy loss density, the gamma rays from deexcitation and neutron capture are generally difficult to detect with tracker detectors.
The deexcited gamma rays also have a certain energy threshold for neutrons.
On the other hand, gamma rays emitted by neutron capture do not require any detection threshold for neutrons.
Neutrons thermalized by proton recoils remain in the liquid scintillator and water Cherenkov detectors for several hundred microseconds.
In these detectors, most of them are captured by protons emitting a 2.2\,MeV gamma ray.
Large liquid scintillator detectors such as KamLAND~\cite{PhysRevD.107.072006} and JUNO~\cite{An_2016} are good at measuring these gamma rays because of their large light yield.
On the other hand, detecting the 2.2\, MeV gamma ray is challenging due to its small light yield for water Cherenkov detectors such as Super-Kamiokande~\cite{Abe_2022}.
Ongoing detector updates are overcoming this issue by dissolving gadolinium (Gd) in water, such as Super-Kamiokande Gadolinium~\cite{ABE2022166248} and ANNIE experiment~\cite{anghel2015letter}.
Neutron capture on Gd has a much larger cross section than protons, carbons, and oxygens.
In addition, it emits gamma rays of 8\,MeV in total.
Therefore, Gd-loaded water Cherekov detectors can detect neutron capture with high efficiency.

These detectors are expected to improve the results of various physics analyses by using the measured neutron multiplicity to enhance flavor identification or signal-to-background ratio.
The following three physics targets are discussed: {\it CP}-phase measurements, diffused supernova neutrino background (DSNB) searches, and nucleon decay searches.
In the {\it CP}-phase measurements, the charged-current quasielastic (CCQE) interaction ($\nu_\mu + n \rightarrow \mu^- + p$ and $\bar{\nu}_\mu + p \rightarrow \mu^+ + n$) is used as a signal.
A comparison of neutrino and antineutrino oscillation is achieved by using accelerator neutrino to switch between neutrino and antineutrino dominant modes.
However, some wrong-sign neutrinos contaminate the beam during the beam production process.
The presence or absence of neutrons in the final state is expected to reduce the contamination in event selections.
In the DSNB searches, the inverse beta decay ($\bar{\nu}_e + p \rightarrow e^{+}+n$) is usually used as a signal, and the dominant background is atmospheric neutrinos.
The background can be drastically reduced by requiring single neutron capture in the event selection while maintaining high signal efficiency~\cite{Harada_2023}.
In the nucleon decay search, the signal usually does not accompany a neutron in the final state to violate the baryon number.
If a nucleon bound in a nucleus decays, the residual nucleus is deexcited and frequently emits neutrons in the final state.
The dominant background is again atmospheric neutrinos.
Therefore, the signal-to-background ratio can be increased by requiring the absence of neutron capture in the event selection~\cite{PhysRevD.102.112011}.
These analyses entirely rely on the neutrino (or nucleon decay) Monte Carlo event generators to predict how the particles emitted by neutrino interactions (or nucleon decays) respond in detectors.
Therefore, the prediction accuracy of neutron multiplicity in the generators determines how much we can improve the physics results.
\par
Returning to residual nuclear deexcitation, the fact that this process was not considered in most of the generators that employ the impulse approximation was a critical problem.
This process was not simulated with few exceptions because of the small energy of the deexcited particles.
However, neutron capture detection is susceptible to deexcitation since no detection threshold is imposed in neutron energies.
Therefore, the highest priority is constructing a comprehensive description and simulator of neutrino-nucleus interactions, including nuclear deexcitation.
NEUT~\cite{Hayato2021} considers the deexcitation process only for $^{16}$O based on Ref.~\cite{kobayashi2006deexcitation}, mainly for gamma rays.
A more precise simulator applicable to $^{16}$O and other nuclei is necessary.
More simulation studies of the deexcitation process for application to neutrino interactions and nucleon decays have been discussed extensively in recent years~\cite{PhysRevC.48.1442,PhysRevD.67.076007,PhysRevD.103.053001,HU2022137183,PhysRevD.108.112008,PhysRevLett.108.052505,PhysRevD.99.012002}.
Notable points in these studies are briefly summarized below:
Deexcitation simulation studies using TALYS~\cite{Koning2023}, similar to this paper, are discussed in Refs.~\cite{PhysRevD.103.053001,HU2022137183}.
However, since these were intended for use with liquid scintillator detector JUNO, these dealt only with carbon target and is a closed source,
making it impossible to use with other detectors.
A simulation study combining a neutrino generator NuWro~\cite{PhysRevC.86.015505} and FSI model INCL coupled with deexcitation code ABLA~\cite{PhysRevC.87.014606} was made focusing on carbon target in Ref.~\cite{PhysRevD.108.112008}.
ABLA is provided as an option to describe the deexcitation process in the INCL code.
Another neutrino generator GENIE~\cite{ANDREOPOULOS201087} can also simulate the deexcitation process using ABLA in principle since INCL is already implemented into it, although a detailed study has yet to emerge.
ABLA lacks predictive power for gamma rays from low-energy discrete excited states, as commented in Ref.~\cite{PhysRevD.108.112008}, and thus may be unsuitable for liquid scintillators and Cherenkov detectors.
Another study using PEANUT in FLUKA~\cite{BATTISTONI201510} was performed by liquid argon time projection chambers, ArgoNeuT~\cite{PhysRevD.99.012002}.
The MeV-scale experimental data of ArgoNeuT shows good agreement with FLUKA prediction.
This result indicates that PEANUT offers an excellent deexcitation model, though more verification in the GeV region and other nuclei, such as carbon and oxygen, is desired.
In addition, incorporating PEANUT into the existing neutrino generators, which has yet to be achieved, might entail technical challenges.
To summarize, there are several relevant studies, but many are limited in applicable nuclei, have limited validation studies, or are closed-source.
The closed sources make their integration into the existing neutrino generators impossible and lead to limited use and validation of the deexcitation simulators in neutrino experiments.
\par
Therefore, the author developed a dedicated software, NucDeEx, to simulate nuclear deexcitation.
Two notable features of NucDeEx are as follows:
It can be applied to $^{12}\text{C}$ and $^{16}\text{O}$ aiming for use in liquid scintillator and water Cherenkov detectors.
Moreover, NucDeEx's code is provided as a package applicable to existing neutrino and nucleon decay Monte Carlo event generators.
The former is because observing particles from deexcitation becomes a problem with these detectors.
The latter is essential to be used in the running and future experiments, a crucial difference from similar studies that are not open source.
This paper is organized as follows:
Sec.~\ref{sec:overview} describes the overall simulation strategy;
Sec.~\ref{sec:results} then presents its validation using other prediction and non-neutrino beam experiments;
Sec.~\ref{sec:application} finally shows the impacts of the deexcitation process on neutron multiplicity associated with neutrino interactions.
In this section, three widely used neutrino Monte Carlo event generators are discussed: NEUT~\cite{Hayato2021}, NuWro~\cite{PhysRevC.86.015505}, and GENIE~\cite{ANDREOPOULOS201087}.
The NucDeEx codes and sample codes used with the generators are available at GitHub~\cite{code}.

\section{Simulation overview} \label{sec:overview}
Neutrino Monte Carlo event generators are event-by-event simulators.
The deexcitation simulator NucDeEx presented in this paper is also built as an event-by-event simulator for use with it.
The basic concepts of a predecessor deexcitation simulator to NucDeEx were explained in Refs.~\cite{Abe_2021,PhysRevD.107.072006}.
The predecessor was composed of two software: TALYS~\cite{Koning2023} and Geant4~\cite{AGOSTINELLI2003250}.
TALYS, an open-source package for nuclear reaction, calculates comprehensive branching ratios based on the Hauser-Feshbach model~\cite{PhysRev.87.366}.
Geant4, a widely used particle simulation package, calculates event-by-event kinematics using the branching ratios from TALYS.
The predecessor was used with NuWro in atmospheric neutrino analysis at KamLAND~\cite{PhysRevD.107.072006}.
The analysis result showed excellent agreement in neutron multiplicity between predictions and observations.
Note that the predecessor is not open-source and integrated into NuWro.
\par
In order to make NucDeEx readily available for other experiments,
it must be inter-operative with the existing neutrino Monte Carlo event generators.
The predecessor requires Geant4, but its large library set may introduce complexity and compatibility issues.
Aiming to minimize dependencies for implementation to the existing generators, the author modified the code to compute kinematics using ROOT~\cite{Brun:1997pa} instead of Geant4.
Since the generators already depend on ROOT, the fact that the deexcitation simulator depends on ROOT will not be a problem.
The NucDeEx code has been opened with various other modifications,
but the idea is similar to previous studies explained in Refs.~\cite{Abe_2021,PhysRevD.107.072006}.

\subsection{Simulation procedure} \label{sec:procedure}
\begin{figure}[htbp]\centering
\includegraphics[width=0.80\columnwidth]{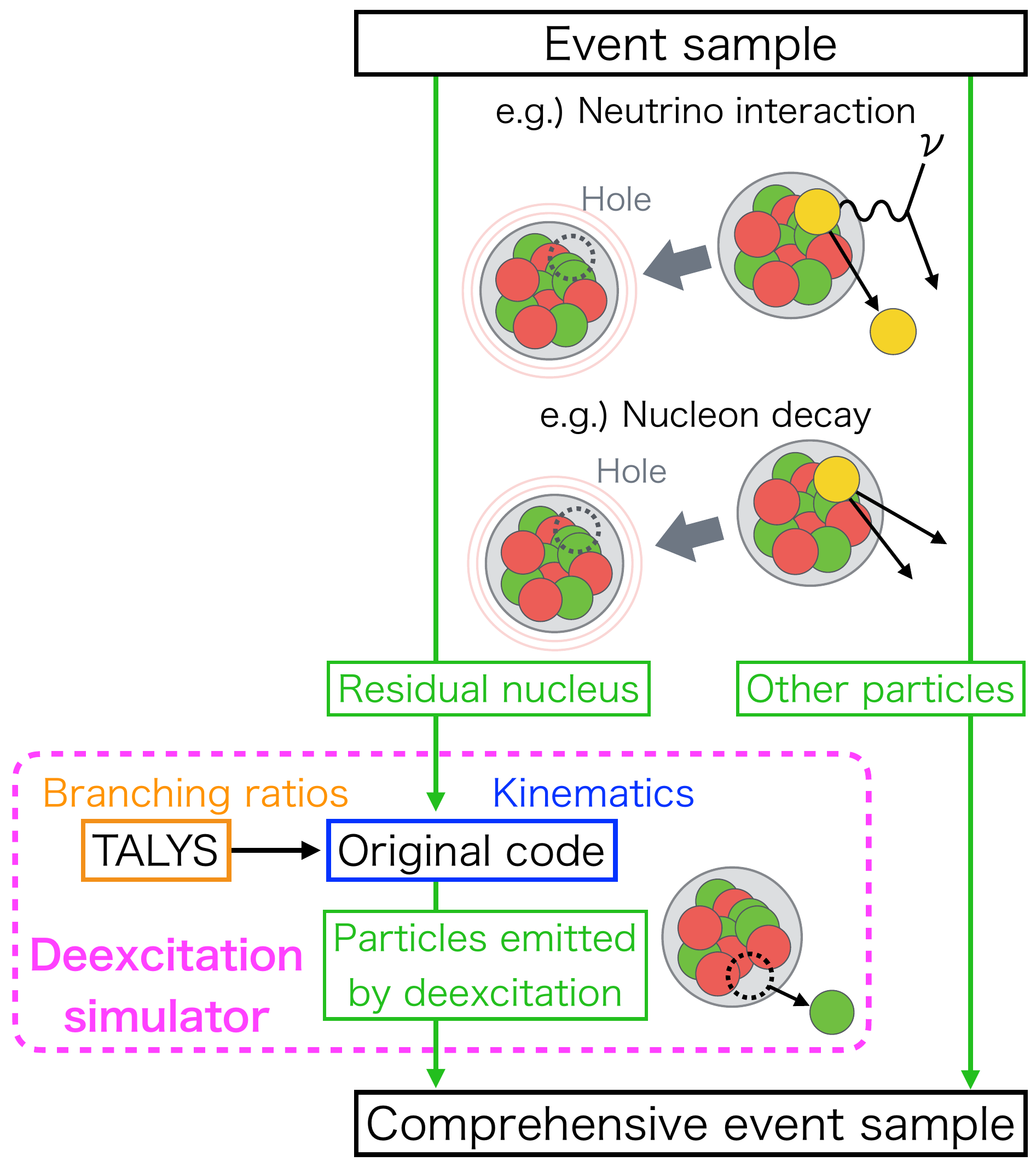}
\caption{A procedure of the deexcitation simulation.
         From the event sample of neutrino interactions or nucleon decays, information on the residual nucleus is extracted and passed to the deexcitation simulator, NucDeEx.
         Using the branching ratios calculated with TALYS~\cite{Koning2023} (denoted as the orange part), particle kinematics is simulated (denoted as the blue part).
         Combining particle information, we get a comprehensive event sample, which includes the deexcitation process.
        }
\label{fig:overview_deex}
\end{figure}

Figure~\ref{fig:overview_deex} shows a procedure of the deexcitation simulation.
The information of the residual nucleus after the FSI is extracted from an event sample of neutrino interactions or nucleon decays.
Particle kinematics is determined by the kinematics simulator using ROOT.
The branching ratios and separation energies necessary to decide the kinematics are calculated with TALYS and tabulated in advance.
Therefore, NucDeEx depends on TALYS regarding simulation outputs but is independent regarding software codes.
The branching ratios and kinematics simulation details are described in Sec.~\ref{sec:br} and Sec.~\ref{sec:kinematics}, respectively.

The key point of this design is that TALYS only calculates branching rations for nuclei,
not simulating particle kinematics event-by-event and not giving excitation energy distribution.
That is why the author prepares the particle simulation using Geant4 or ROOT.
We need to know nuclear species, hole state, and excitation energy.
These parameters are extracted from the outputs of neutrino or nucleon decay Monte Carlo event generators.

\subsection{Branching ratios for \mbox{\boldmath $p_{3/2}$}-hole states} \label{sec:br_p}
Due to its low excitation energy, the deexcitation process from $p_{3/2}$-hole states is simple.
The branching ratios for $p_{3/2}$-hole states were measured or predicted well in Refs.~\cite{PANIN2016204,PhysRevC.48.1442}.
NucDeEx refers to these data instead of TALYS.
\par
In the simple shell model picture of $^{12}$C, two nucleons lie in the $s_{1/2}$ shell, and four nucleons lie in the $p_{3/2}$ shell.
No nucleon exists in the $p_{1/2}$ shell, which lies a few MeV above the $p_{3/2}$ shell.
In this case, the $p_{3/2}$-hole state always goes to the ground state (g.s.); no deexcitation happens.
However, according to more precise shell model calculations,
the $p_{1/2}$ shell is partially filled with nucleons at a certain probability due to nucleon-nucleon correlation.
As a result of this paring effect, about $20\pm5$\% of the $p_{3/2}$-hole states are expected to have excitation energies of a few MeV~\cite{PhysRevD.67.076007}.
Table.~\ref{tab:br_p_12C} shows a summary of the branching ratios for $p_{3/2}$-hole states of $^{11}\text{B}^{*}$ and $^{11}\text{C}^{*}$ used in NucDeEx.
The experimental data for $^{11}\text{B}^*$ shown in this table agrees well with the predicted value mentioned above.
Since no experimental data for $^{11}\text{C}^*$ could be found, an analogy was assumed:
While the same branching ratio is used, the excitation states are changed to ones that have the same spin-parity and similar excitation energy.
The shell structures of neutrons and protons are identical except for the Coulomb potential of about 2-3\,MeV.
Especially in the low excitation energy region, there are excited states with the spin-parity at close energies for $^{11}\text{B}^*$ and $^{11}\text{C}^*$.
This fact suggests, to some extent, the validity of the analogy of assuming isospin symmetry.
In reality, the symmetry is broken by the Coulomb potential, and the excitation energies are slightly different.
Therefore, experiments on $^{11}\text{C}^*$ to evaluate the effect of the symmetry breaking are desired.
\par
As for $^{16}$O, all shells including $p_{1/2}$ are filled with nucleons.
The $p_{1/2}$-hole states always go to the g.s. and $p_{3/2}$-hole states go to excited states.
Table.~\ref{tab:br_p_16O} shows summaries of the branching ratios for $p_{3/2}$-hole states of $^{15}\text{N}^{*}$ and $^{15}\text{O}^{*}$ used in NucDeEx.
It sometimes transitions to the excited states that emit a proton or multiple gamma rays.
As well as $^{11}\text{C}^*$, no data for $^{15}\text{O}^*$ could be found, so the same analogy is assumed.
\par
From the $p_{3/2}$-hole states, neither oxygen nor carbon is accompanied by a deexcited neutron.
However,  gamma rays affect the visible energy in neutrino detectors.
The contributions from the gamma ray become relatively large and require prediction accuracy, especially in DSNB searches looking at low energy regions below 30\,MeV~\cite{PhysREvD.104.122002} and proton decay searches~\cite{PhysRevD.102.112011}.
The probability of occurrence of $p_{3/2}$-hole state is about twice larger than that of $s_{1/2}$-hole states, which is discussed in Sec.~\ref{sec:br}.
NucDeEx precisely considers the deexcitation from $p_{3/2}$-hole states, although they do not affect neutron multiplicity.

\begin{table}[htbp] \centering
\caption{Excited states and branching ratios for $p_{3/2}$-hole states of $^{11}\text{B}^{*}$ and $^{11}\text{C}^{*}$ used in NucDeEx.
         The excitation energy and gamma ray energy are denoted as $E_x$ and $E_\gamma$, respectively.
         The branching ratios (Br) for $^{11}\text{B}^{*}$ are measured by Panin {\it et al.}~\cite{PANIN2016204}.
         All gamma rays emitted from these excited states are single.
         }
\label{tab:br_p_12C}
\begin{tabular*}{0.9\columnwidth}{@{\extracolsep{\fill}}lccccc} \hline \hline
Nucleus & $E_x$\,(MeV) & $J^\pi$ & Br & $E_\gamma$\,(MeV) \\ \hline
$^{11}\text{B}^*$& 0 (g.s.) & $3/2^{-}$ & 0.82 & - \\
                 & 2.13    & $1/2^{-}$ & 0.10 & 2.13  \\
                 & 5.02    & $3/2^{-}$ & 0.08 & 5.02 \\  \hline
$^{11}\text{C}^*$& 0 (g.s.) & $3/2^{-}$ & 0.82 & - \\
                 & 2.00    & $1/2^{-}$ & 0.10 & 2.00 \\
                 & 4.80    & $3/2^{-}$ & 0.08 & 4.80 \\ \hline \hline
\end{tabular*}
\end{table}

\begin{table}[htbp] \centering
\caption{Excited states and branching ratios for $p_{3/2}$-hole states of $^{15}\text{N}^{*}$ and $^{15}\text{O}^{*}$ used in NucDeEx.
         The excitation energy and gamma ray energy are denoted as $E_x$ and $E_\gamma$, respectively.
         The branching ratios (Br) for $^{15}\text{N}^{*}$ are from~\cite{PhysRevC.48.1442}.
         Gamma rays emitted from these excited states can be multiple.
         The relative branching ratios for gamma rays (RBr$_\gamma$) are based on TALYS~\cite{Koning2023}.
         Only those with RBr$_\gamma>0.01$ are listed.
         From excited states with high excitation energies,
         a proton is emitted with 100\%, having a energy written as $E_p$.
         }
\label{tab:br_p_16O}
\begin{tabular*}{1.0\columnwidth}{@{\extracolsep{\fill}}lcccccc} \hline \hline
Nucleus & $E_x$\,(MeV) & $J^\pi$ & Br & RBr$_\gamma$ & $E_\gamma$\,(MeV) & $E_p$\,(MeV)\\ \hline
$^{15}\text{N}^*$& 6.32 & $3/2^-$ & 0.87 & 1.00 & 6.32 & - \\
                 & 9.93 & $3/2^-$ & 0.06 & 0.02 & $2.63 + 7.30$ & -  \\
                 &      &         &      & 0.04 & $3.61 + 6.32$ & -  \\
                 &      &         &      & 0.13 & $4.63 + 5.30$ & -  \\
                 &      &         &      & 0.13 & $4.66 + 5.27$ & -  \\
                 &      &         &      & 0.65 & $9.93$ & -  \\
                 & 10.70 & $3/2^-$ & 0.06 & - & - & 0.5 \\ \hline
$^{15}\text{O}^*$& 6.18 & $3/2^-$ & 0.87 & 1.00 & 6.18 & - \\
                 & 9.61  & $3/2^-$ & 0.06 & - & - & 2.31 \\
                 & 10.48 & $3/2^-$ & 0.06 & - & - & 3.18\\ \hline \hline
\end{tabular*}
\end{table}

\subsection{Calculation of branching ratios for \mbox{\boldmath $s_{1/2}$}-hole and multi-nucleon hole states using TALYS} \label{sec:br}
In the case of high excitation energies, such as $s_{1/2}$-hole and multi-nucleon hole states, the deexcitation process becomes complicated and various particles are emitted.
Branching ratios for $\gamma$, $\alpha$, $n$, $p$, deuteron ($d$), triton ($t$), and $^{3}$He are calculated with TALYS version 1.96.
This calculation corresponds to the orange part in Fig.~\ref{fig:overview_deex}.
A deexcitation simulation study using SMOKER code~\cite{COWAN1991267} was made in Ref.~\cite{PhysRevD.67.076007}.
SMOKER is another nuclear simulator that uses the Hauser-Feshbach model like TALYS, but it does not deal with deuteron, triton, and $^{3}$He emissions.
One of the features of TALYS is its ability to describe deexcitation, even for heavy ions.
TALYS provides a global optical model potential to determine the transmission coefficients and several sophisticated level density models~\cite{KONING200813}.
The author configured TALYS to calculate the optical model for any compound nucleus.
The default level density model is a combination of the constant temperature and a Fermi gas, which was adopted in previous studies~\cite{Abe_2021,PhysRevD.107.072006}.
The author found that another model, the back-shifted Fermi gas model~\cite{DILG1973269} gave better agreement with experimental data shown in Sec.~\ref{sec:results}; therefore, the model is adopted in NucDeEx.
\par
\begin{figure*}[htbp]
\begin{minipage}[tl]{0.45\textwidth} \centering
\includegraphics[width=0.95\textwidth]{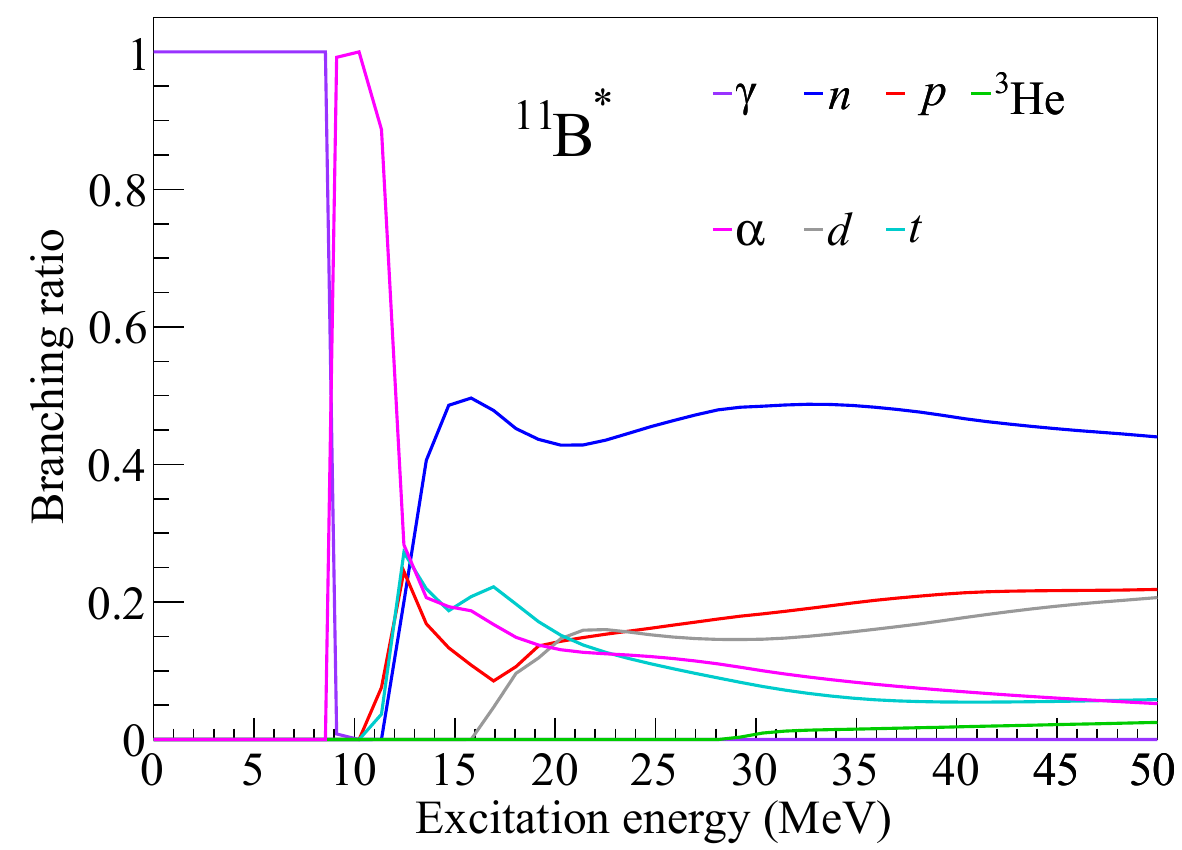}
\end{minipage}
\hspace{0.01\textwidth}
\begin{minipage}[tr]{0.45\textwidth} \centering
\includegraphics[width=0.95\textwidth]{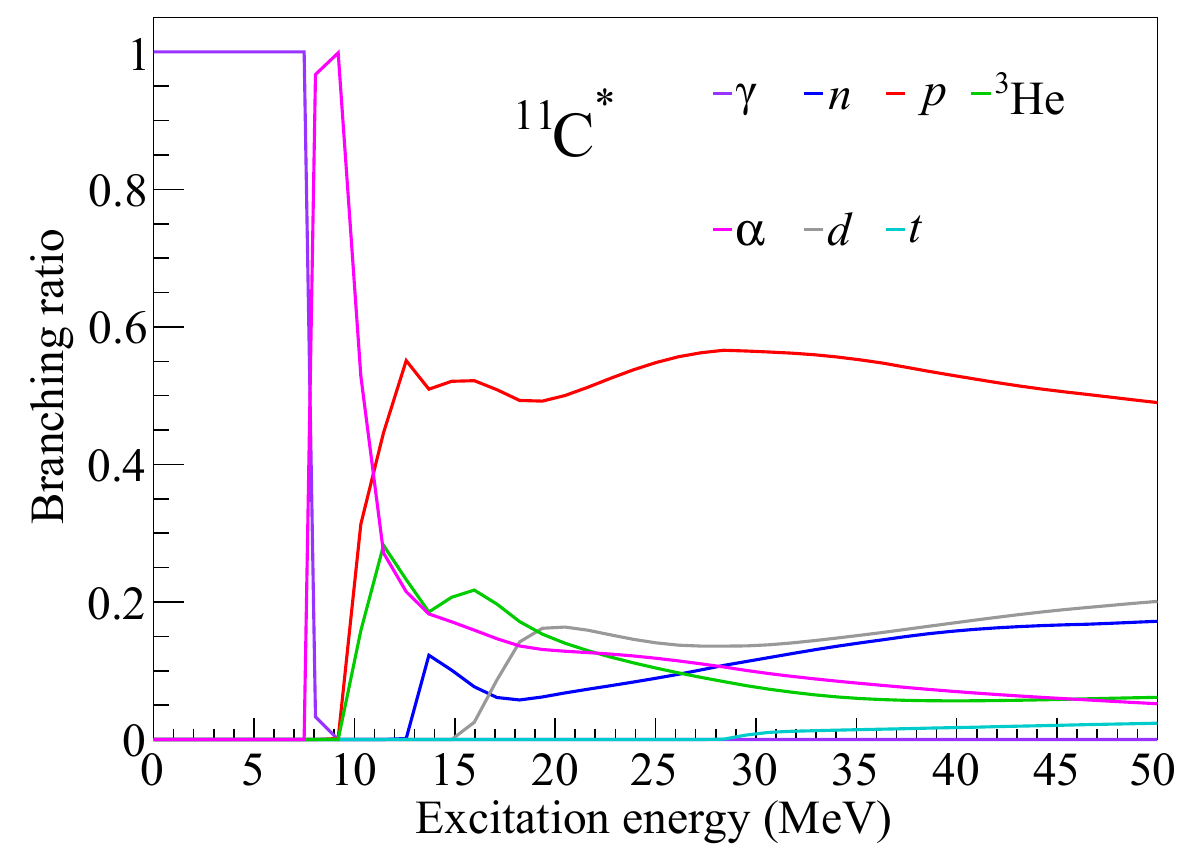}
\end{minipage}
\\
\vspace{10pt}
\begin{minipage}[bl]{0.45\textwidth} \centering
\includegraphics[width=0.95\textwidth]{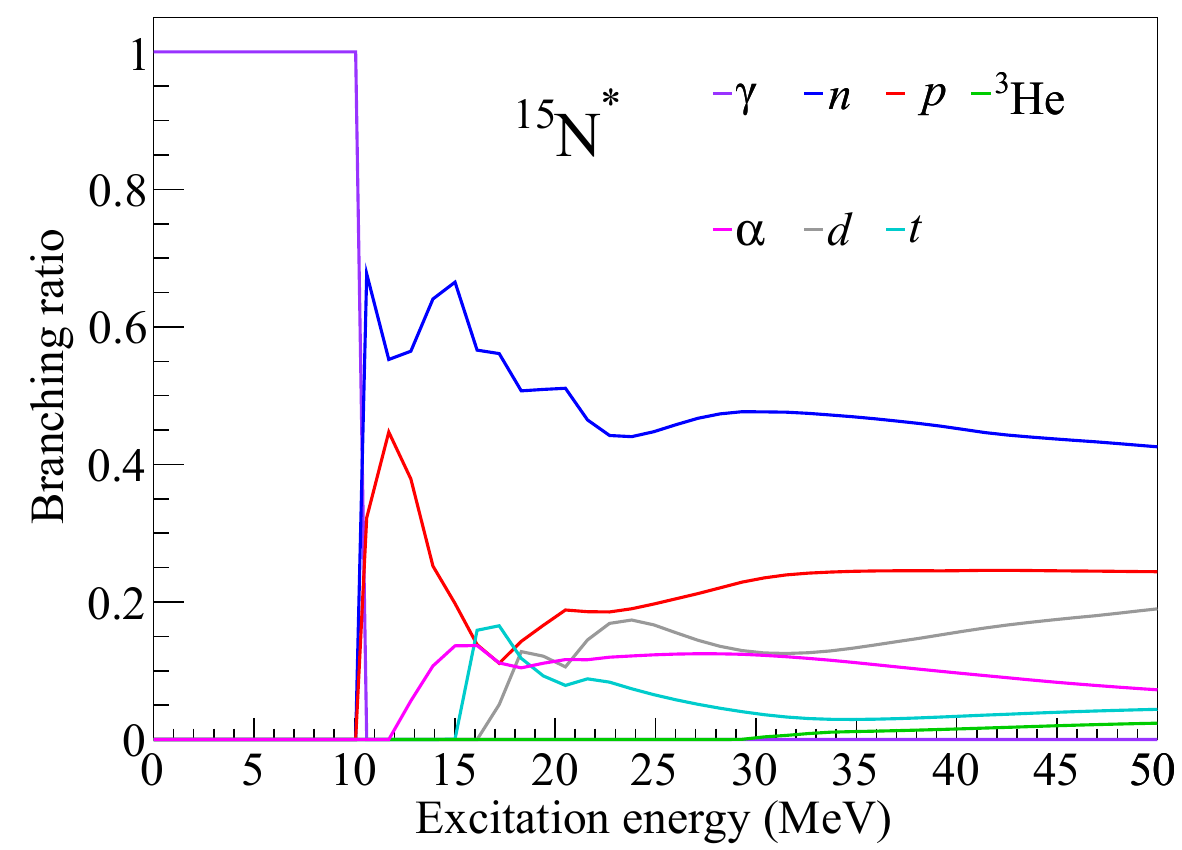}
\end{minipage}
\hspace{0.01\textwidth}
\begin{minipage}[br]{0.45\textwidth} \centering
\includegraphics[width=0.95\textwidth]{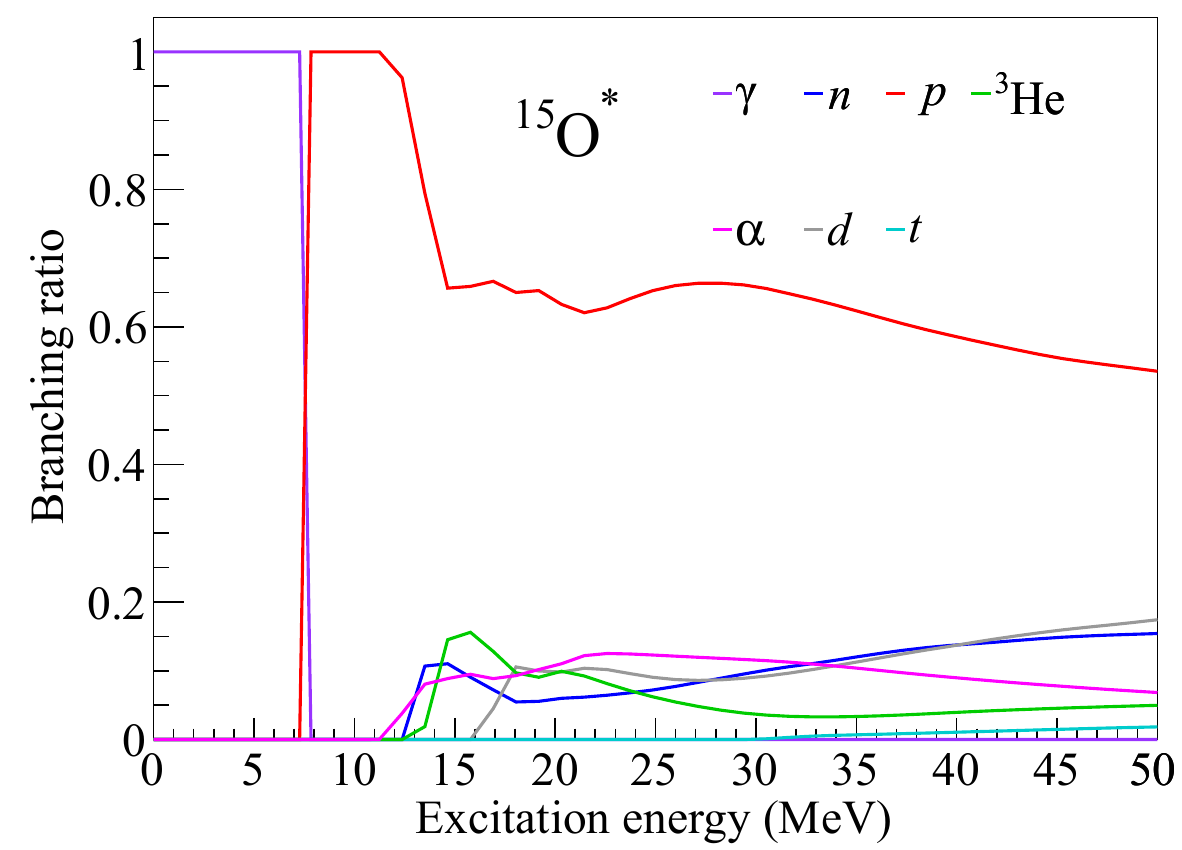}
\end{minipage}
\caption{Branching ratios of $^{11}$B$^*$, $^{11}$C$^*$, $^{15}$N$^*$, and $^{15}$O$^*$ as a function of excitation energy calculated with TALYS~\cite{Koning2023}.
         The spin-parity $J^{\pi}=1/2^{+}$ is set for $s_{1/2}$-hole states.
         }
\label{fig:br_talys}
\end{figure*}

The calculated branching ratios of $s_{1/2}$-hole states from $^{12}$C and $^{16}$O are shown in Fig.~\ref{fig:br_talys}.
The spin parity $J^{\pi}=1/2^+$ is set for application to these states.
At a typical excitation energy of 23\,MeV (30\,MeV) for $^{11}$B$^*$ ($^{15}$N$^*$),
neutron emission dominates about 50\%.
The results indicate that the deexcitation process has a non-negligible impact on neutron multiplicity.
Comparing the results of $^{11}\text{B}^*$ and $^{11}\text{C}^*$, related to the $^{12}$C target, there is an approximate isospin-symmetry in the branching ratios:
For example, the branching ratio of $n$ for $^{11}\text{B}^*$ is similar to one of $p$ for $^{11}\text{C}^*$.
On the contrary, the symmetry is almost broken for $^{15}\text{N}^*$ and $^{15}\text{O}^*$ due to their larger Coulomb potential.
A general trend, the branching ratios of $p$ for neutron-hole nuclei are larger than ones of $n$ for proton-hole nuclei, is seen through the results.
In the simulation process, the emitted particles are first determined using the branching ratio shown in Fig.~\ref{fig:br_talys}.
The particle's energy depends on the daughter nucleus's excitation energy to be transited.
This simulation step is determined as explained below.

\begin{figure*}[htbp] \centering
\includegraphics[width=1.0\textwidth]{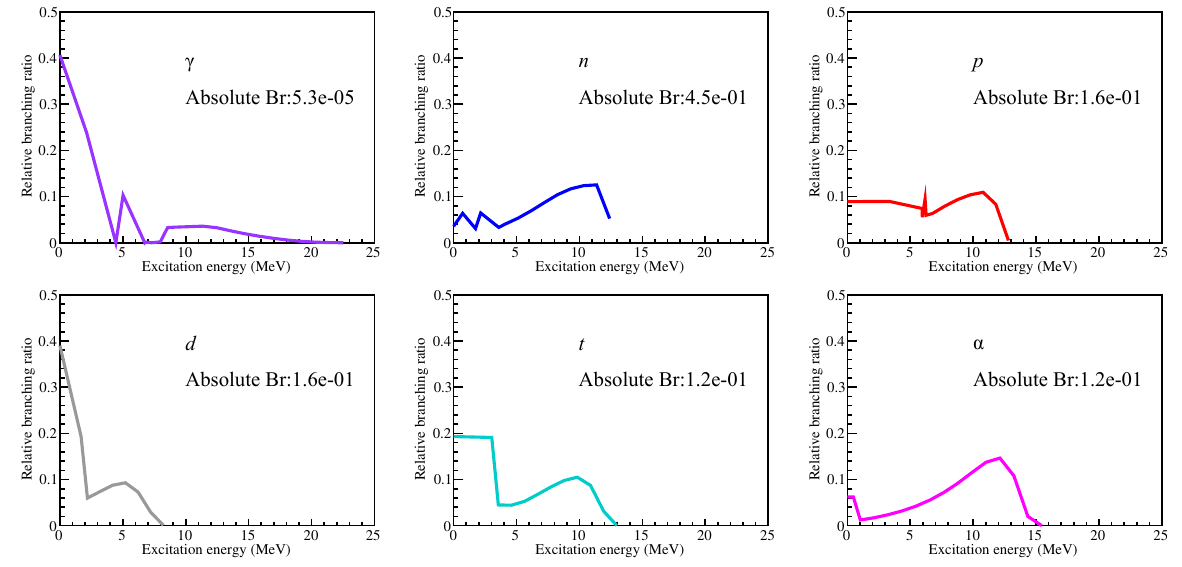}
\caption{Relative branching ratios of $^{11}$B$^*$ with excitation energy of 23.6\,MeV and $J^\pi=1/2^+$ calculated with TALYS~\cite{Koning2023}.
         The text in each panel denotes the absolute branching ratio for each particle emission,
         and the colored lines represent the relative branching ratio as a function of the excitation energy of the corresponding daughter nucleus.
        }
\label{fig:11B_br_Ex}
\end{figure*}

When nuclei have high excitation energy, they rarely go to the ground state by a single-step decay but rather transition to an excited state and decay sequentially.
Thus, we have to calculate the transition probabilities of all possible excited states of the daughter nuclei and the branching ratios from the states.
The complete sets of information are also calculated and tabulated using TALYS.
Figure~\ref{fig:11B_br_Ex} shows an example of the relative branching ratio of each particle as a function of the excitation energy of the corresponding daughter nucleus.
After determining the emitted particles in the previous simulation step, the excited state of the daughter nucleus to be transited is determined using the relative branching ratio shown in Fig.~\ref{fig:11B_br_Ex}.
The relative branching ratios can also be interpreted as probability densities.
The energies of the emitted particles are determined by this simulation step.
NucDeEx repeats these steps until the excitation energy of the daughter nucleus becomes zero.
The branching ratios for all possible daughter nuclei (e.g., $^{10}\text{B}^{*}$ produced via $^{11}\text{B}^*\rightarrow n+^{10}\text{B}^*$) required for these simulation steps are also calculated and tabulated in the same way.
The details are omitted in this paper due to the large data size.

So far, we have discussed single nucleon hole states where we know the spin-parity exactly.
However, multi-nucleon holes are produced by FSI with a probability of about 30\% in GeV-order neutrino reactions.
In the case of such multi-nucleon hole states, most neutrino Monte Carlo generators cannot determine the spin-parity of the residual nucleus because the Fermi gas model used in FSI does not consider shell levels.
Therefore, branching ratios of multi-nucleon hole states are calculated using TALYS without specifying the spin-parity.
The spin-parity is determined according to the level density model employed in TALYS, the back-shifted Fermi gas model.

\subsection{Kinematics simulation} \label{sec:kinematics}
The kinematics simulation, corresponding to the blue part in Fig.~\ref{fig:overview_deex}, depends on ROOT libraries.
It considers the separation energies and the four-dimensional momentum conservation.
The deexcited particles are assumed to be emitted isotropically in the center-of-mass frame, namely the rest frame of the mother nucleus.
The deexcitation decay process could be boosted because the residual nuclei tend to have about 200\,MeV/c in momentum.
The Lorentz boost from the center-of-mass frame to the laboratory frame is also considered in this simulation.
The effect of the Lorentz boost is not so large but introduced to accurately reproduce the detection thresholds' effect when compared with experimental data in Sec.~\ref{sec:results}.

\section{Results of \mbox{\boldmath $s_{1/2}$}-hole state simulation} \label{sec:results}
Before showing the results of NucDeEx with neutrino interactions, its performance is discussed in this section.
In order to validate the performance independently of neutrino interactions and generators,
the deexcitation process from specific nuclear states is compared with experimental data and other predictions.
We focus on deexcitation from $s_{1/2}$-hole states of $^{11}$B$^*$ and $^{15}$N$^*$.
Two experiments are used for comparisons:
Panin {\it et al.} measured $^{12}\text{C}(p,2p)^{11}\text{B}^*$ reaction and associated deexcitation particles using $^{12}$C beam at an energy of about 400\,MeV~\cite{PANIN2016204}.
Single-step decays of $n$, $d$, and $\alpha$ were measured in this experiment.
Yosoi {\it et al.} measured $^{12}\text{C}(p,2p)^{11}\text{B}^*$ and $^{16}\text{O}(p,2p)^{15}\text{N}^*$ reactions using proton beam at an energy of 392\,MeV~\cite{YOSOI2003255,Yosoi2004,kobayashi2006deexcitation}.
They measured single-step and multistep decays of $n$ (only for $^{15}\text{N}^*$), $p$, $d$, $t$ and $\alpha$.
The branching ratios of $^{3}$He are negligibly small for these nuclei.
Since both experiments have high incident energies, the impulse approximation is applicable, i.e., the types of incident particles do not affect the deexcitation process.
Note that the single-step (multistep) decays in this paper are identical to the two-body decay (three-body decay and sequential decay) in Refs.~\cite{YOSOI2003255,Yosoi2004}.
\par
\begin{figure}[htbp]\centering
\includegraphics[width=0.80\columnwidth]{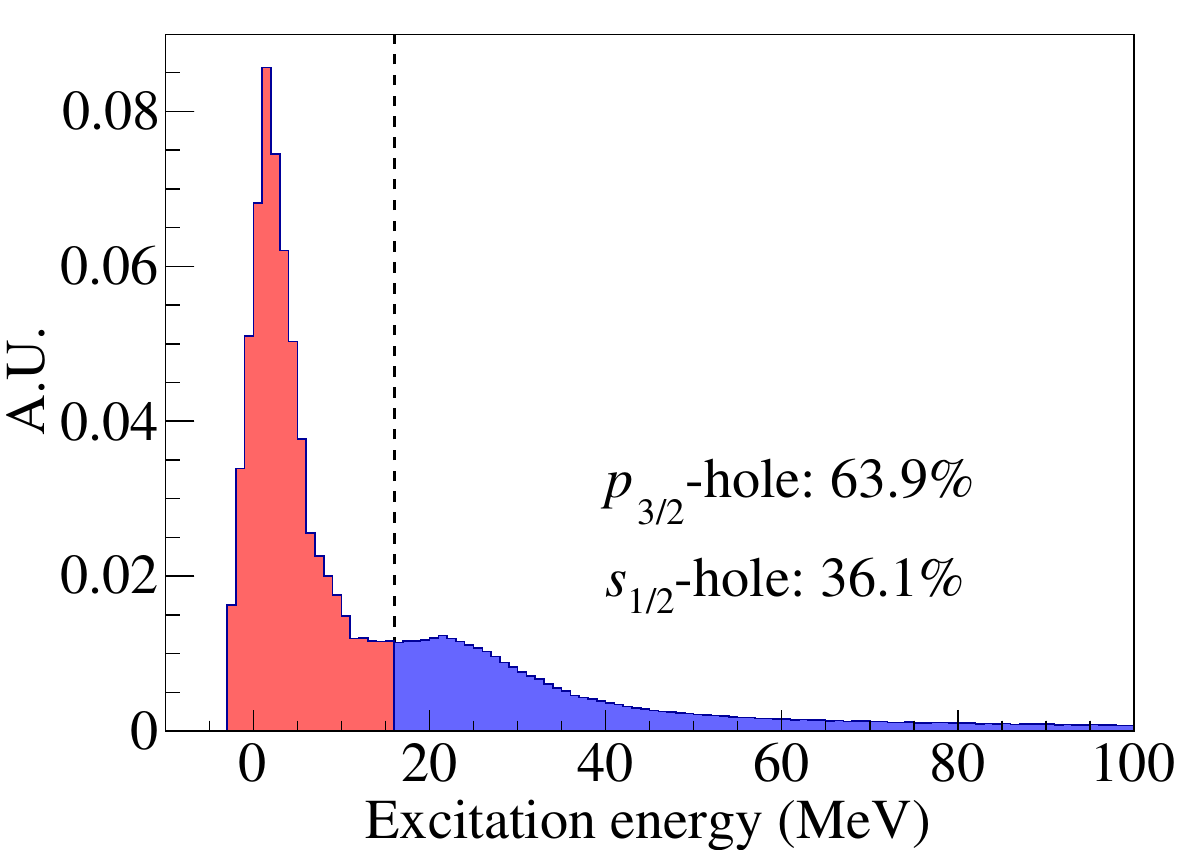}
\\
\vspace{5pt}
\includegraphics[width=0.80\columnwidth]{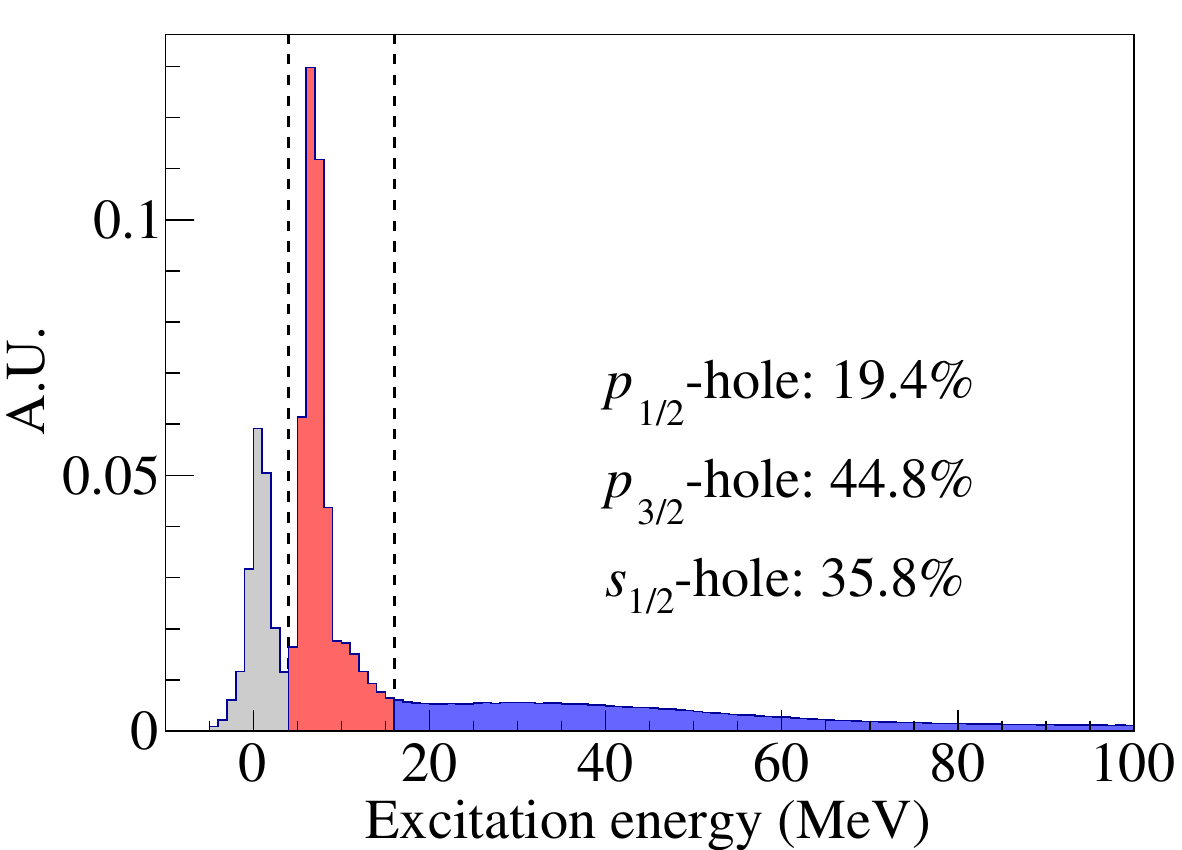}
\caption{Excitation energy distribution for $^{11}\text{B}^*$ (top) and $^{15}\text{N}^*$ (bottom) based on the Benhar SF~\cite{PhysRevD.72.053005}.
         The hole states are determined according to Table~\ref{tab:Ex_range}.
         The blue, red, and gray histograms denote contribution from $s_{1/2}$-, $p_{3/2}$-, and $p_{1/2}$-hole states, respectively.
         The values in panels represent the probability of each hole state.
        }
\label{fig:Ex}
\end{figure}

The excitation energy must be specified instead of the neutrino Monte Carlo event generators in the comparisons.
It is determined according to the spectral function (SF) by Benhar {\it et al.}~\cite{PhysRevD.72.053005}, which is commonly used in generators, shown in Fig.~\ref{fig:Ex}.
The SF provides probability distribution as a function of removal energy.
The excitation energy is obtained by subtracting the separation energy.
As shown in Fig.~\ref{fig:Ex}, this subtraction causes nonphysical negative excitation energies.
We need more high-precision SF data considering discrete excited states to overcome this issue.
There are two or three peaks in the excitation energy distribution, corresponding to $s_{1/2}$-, $p_{3/2}$-, and $p_{1/2}$-hole states.
In the following comparisons, excitation energies above 16 or 20\,MeV and below 35 or 40\,MeV corresponding to the $s_{1/2}$-hole states are selected.

\subsection{Deexcitation from $^{11}$B$^*$} \label{sec:results_11B}
\begin{figure}[htbp] \centering
\includegraphics[width=1.0\columnwidth]{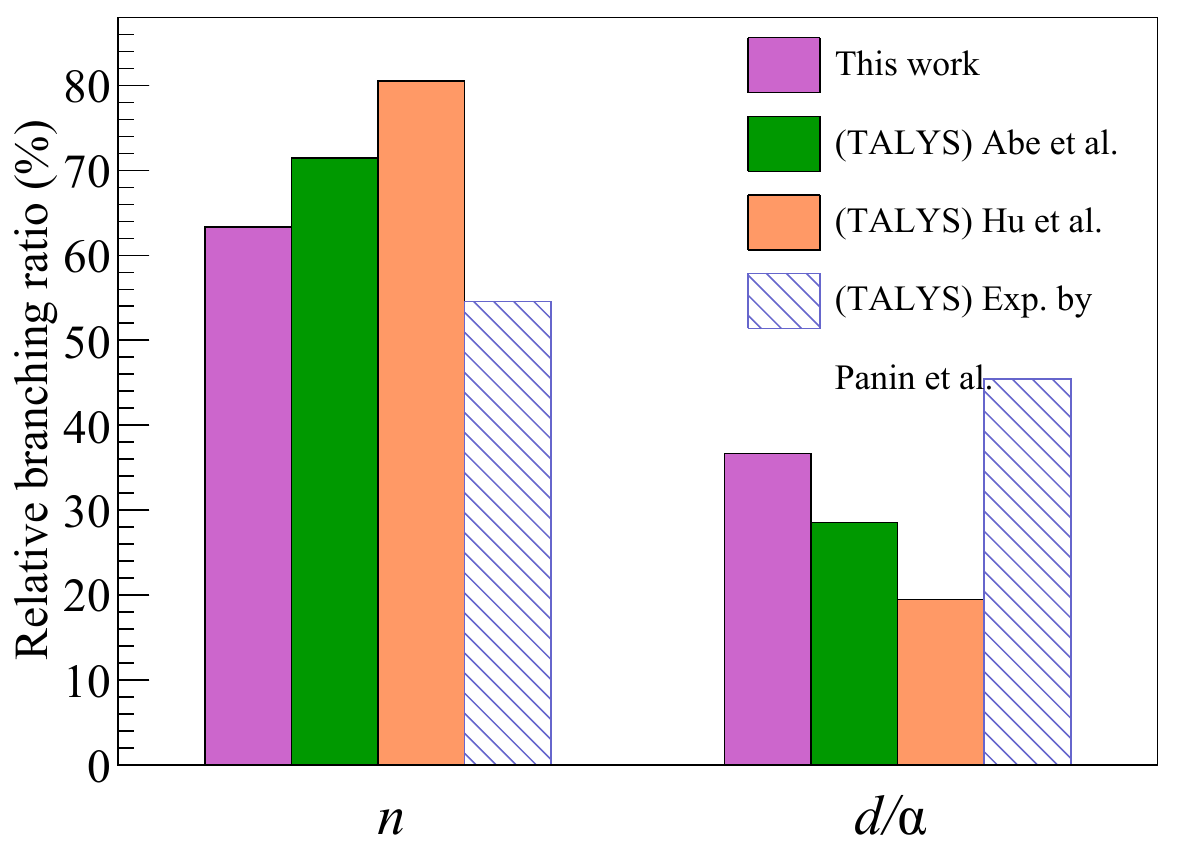}
\caption{Comparison of relative branching ratios of $n$ and $d/\alpha$ for $^{11}$B$^*$ with excitation energies of 16-35\,MeV.
         The magenta histograms show the results using NucDeEx,
         the green histograms show the previous study from~\cite{PhysRevD.107.072006},
         and the orange histograms show the predicted results by Hu {\it et al.} using TALYS~\cite{HU2022137183}.
         The experimental data by Panin {\it et al.}~\cite{PANIN2016204} is shown in the blue-hatched histograms.
         No experimental uncertainty is available.
         The branching ratios shown here account for only single-step decays.
        }
\label{fig:11B_nda}
\end{figure}

A comparison with the $^{12}$C beam experiment of relative branching ratios for $^{11}\text{B}^*$ with 16-35\,MeV excitation energy range is shown in Fig.~\ref{fig:11B_nda}.
All predictions shown here are based on TALYS.
The differences come from the excitation energy distribution and level density model adopted in TALYS.
The excitation energy distributions assumed in these predictions are different and slightly change the branching ratios:
This work and results from Hu {\it et al.} used the SF, while the previous study (the green histogram) assumed the Lorentzian distribution with 23\,MeV as the mean and 14\,MeV as the FWHM.
The level density model is the leading cause of the differences.
As mentioned in Sec.~\ref{sec:br}, NucDeEx uses the back-shifted Fermi gas model, but the others use another level density model: A combination of the constant temperature and a Fermi gas model.
The result of this work gives the best agreement with the experiment with uncertainty of about 15\%.
Since this experiment used $^{12}$C beam, the center-of-mass frame was boosted, and particles emitted by deexcitation could be measured at very low thresholds in the laboratory frame.
The agreement between the experimental data and this work indicates that NucDeEx predicts the branching ratios to low energies well.

\begin{figure}[htbp] \centering
\includegraphics[width=1.0\columnwidth]{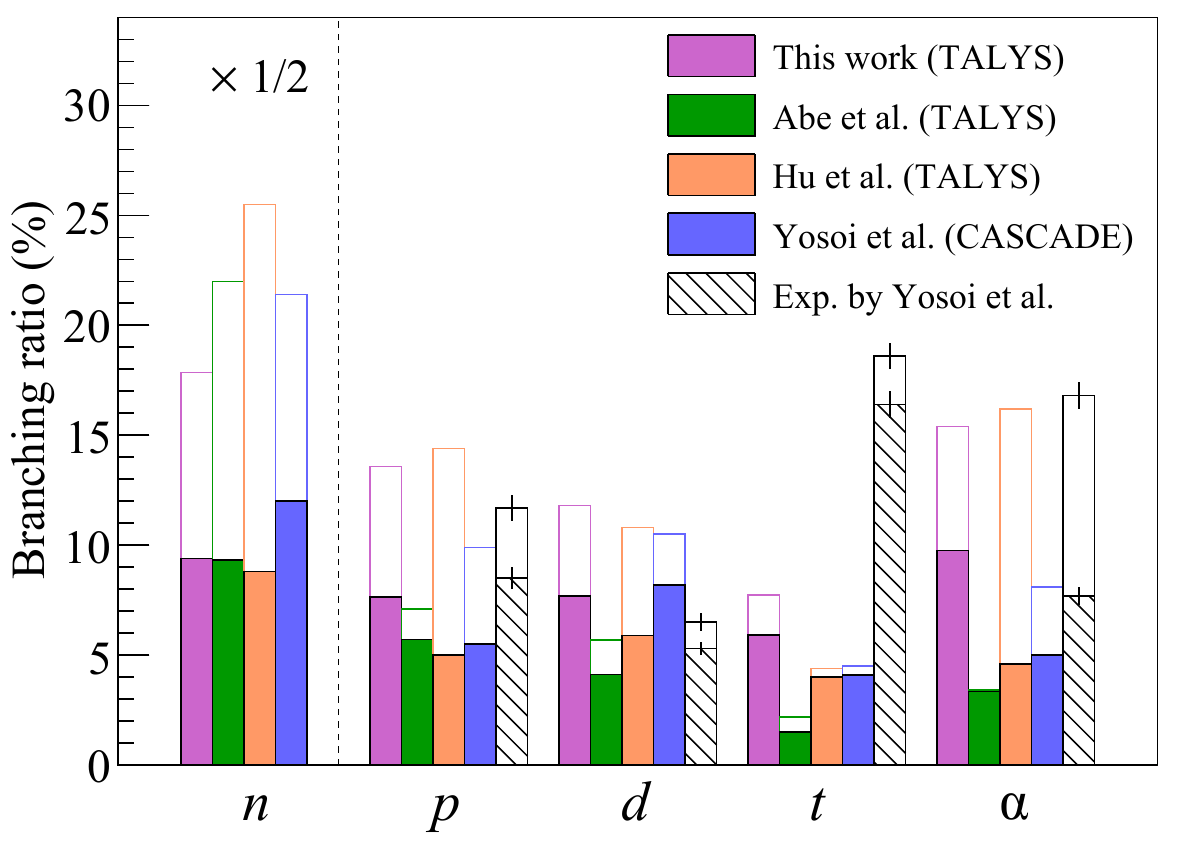}
\caption{Comparison of measured and predicted branching ratios of $n$, $p$, $d$, $t$, and $\alpha$ for $^{11}$B$^*$ with 16-35\,MeV excitation energy.
         The branching ratios of $n$ are scaled by a factor of 1/2.
         The magenta histograms show the results using NucDeEx,
         the green histograms show the previous study from~\cite{PhysRevD.107.072006},
         and the orange histograms show the predicted results from Hu {\it et al.} using TALYS~\cite{HU2022137183}.
         The black histograms denote experimental data from Yosoi {\it et al.} with statistical errors,
         and the authors provide the prediction using CASCADE code written with the blue histograms~\cite{YOSOI2003255}.
         The hatched or filled histograms represent the branching ratios for single-step decays,
         and the open histograms represent those for multistep decays.
         }
\label{fig:11B_br}
\end{figure}

Another comparison with the proton beam experiment of the absolute branching ratios with the same excitation energy range is shown in Fig.~\ref{fig:11B_br}.
We can compare branching ratios of both single-step and multistep decays from this data.
The prediction accuracy of these decays depends not only on the branching ratios but also on the transition probabilities to the excited state, as shown in Fig.~\ref{fig:11B_br_Ex}.
A large discrepancy in $t$ branching ratio is visible.
The back-shifted Fermi gas model gives better agreement but does not solve the issue entirely; a notable difference remains.
The authors of the experiment also discussed this issue, but their simulation (the blue histograms shown in Fig.~\ref{fig:11B_br}) did not reproduce the data either.
This discrepancy of $t$ branching ratio is still an open question to be addressed in the future by performing validation experiments.
The experiment applied the detection thresholds of 3.1-4.6\,MeV depending on particles~\cite{YOSOI2003255}, and these effects are considered in this simulation.
According to NucDeEx prediction, these thresholds lead to considerable inefficiency, about 50\%.
To reduce this inefficiency and to better understand deexcited particles, the ion beam experiment such as Ref.~\cite{PANIN2016204} is preferable.

\subsection{Deexcitation from $^{15}$N$^*$} \label{sec:resutls_15N}
\begin{figure}[htbp] \centering
\includegraphics[width=1.0\columnwidth]{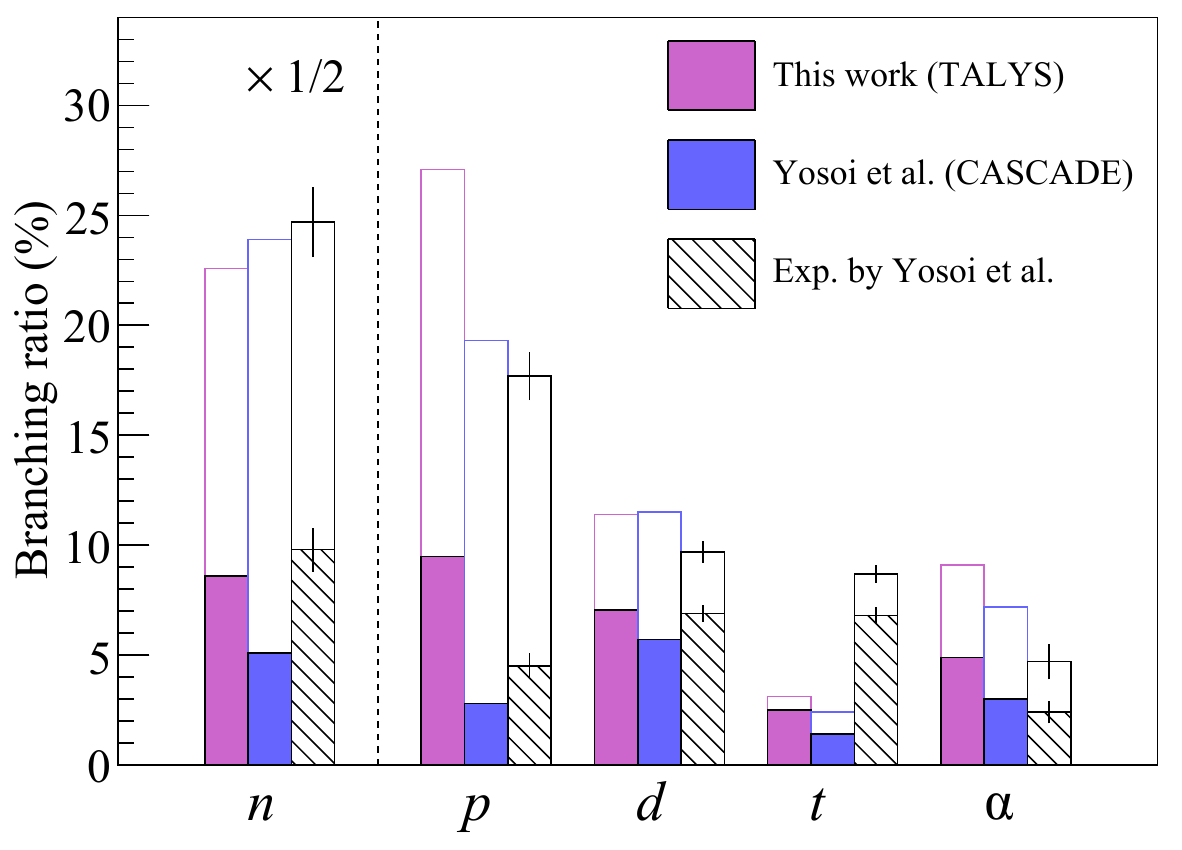}
\caption{Comparison of measured and predicted branching ratios of $n$, $p$, $d$, $t$, and $\alpha$ for $^{15}$N$^*$ with 20-40\,MeV excitation energy.
         The branching ratios of $n$ are scaled by a factor of 1/2.
         The magenta histograms show the results using NucDeEx,
         The black histograms denote experimental data from Yosoi {\it et al.} with statistical errors,
         and the authors provide the prediction using CASCADE code written with the blue histograms~\cite{Yosoi2004}.
         The hatched or filled histograms represent the branching ratios for single-step decays,
         and the open histograms represent those for multistep decays.
        }
\label{fig:15N_br}
\end{figure}

A comparison with the proton beam experiment of the branching ratios for $^{15}\text{N}^*$ with 20-40\,MeV excitation energy range is shown in Fig.~\ref{fig:15N_br}.
Note that this experimental data was taken using the same beamline shown in Fig.~\ref{fig:11B_br}, and the detection thresholds are also considered in this simulation.
Owing to the efforts of installing neutron detectors, the experimental data of $n$ branching ratios are available.
The result of this work agrees with the experimental data in $n$ branching ratio both of single-step and multistep decays within about 20\%.
Together with the results of $n$ branching ratios for $^{11}\text{B}^*$, NucDeEx predictions show reasonable agreements with the experimental datasets.
As well as $^{11}\text{B}^*$, there are large discrepancies for other particles, especially $t$.
Since all predictions do not agree with the experimental data well and these experiments share the same beam line and detectors,
validation experiments are desired.
Conducting a new experiment using $^{16}$O beam is also useful to measure particles with low energy thresholds,
as demonstrated by Panin {\it et al.} introduced in Sec.~\ref{sec:results_11B}.
\par
Kobayashi {\it et al.} made an analysis focusing on the gamma ray from $^{15}\text{N}^*$ using the same experimental data shown in Fig.~\ref{fig:15N_br}~\cite{kobayashi2006deexcitation}.
The branching ratios for gamma rays above 3\,MeV, which are detectable in water Cherenkov detectors, were investigated.
Since multiple gamma rays can be emitted from these states, this measurement was of the total energy released by gamma rays $E_{\gamma,\text{tot}}$.
Table~\ref{tab:15N_gamma} compares gamma ray branching ratios between the experiment and a prediction using NucDeEx.
The prediction agrees with the experiment for low-energy gamma rays below 6\,MeV, but not above 6\,MeV.
The cause of this discrepancy is still under study because gamma rays above several MeV are important in searching for DSNB and nucleon decay using Super-Kamiokande.

\begin{table}[htbp] \centering
\caption{Comparison of branching ratios of $\gamma$ with total energy of 3\,MeV$ < E_{\gamma,\text{tot}} < 6$\,MeV and $E_{\gamma,\text{tot}}>6$ MeV.
         The experimental data from Kobayashi {\it et al.} was measured using proton beams~\cite{kobayashi2006deexcitation}.
         Excitation energies of 16-40\,MeV are selected in this comparison.
         }
\label{tab:15N_gamma}
\begin{tabular*}{1.0\columnwidth}{@{\extracolsep{\fill}}ccc} \hline \hline
 & \multicolumn{2}{c}{Banching ratio (\%)} \\
 & 3\,MeV$ < E_{\gamma,\text{tot}} < 6$\,MeV & $E_{\gamma,\text{tot}} >6$ MeV \\ \hline
Kobayashi {\it et al.} & $27.9\pm1.5^{+3.4}_{-2.6}$ & $15.6\pm1.3^{+0.6}_{-1.0}$ \\
This work & 31.0 & 8.4 \\ \hline \hline
\end{tabular*}
\end{table}

\section{Application to neutrino interactions} \label{sec:application}
In this section, NucDeEx is evaluated with the existing neutrino Monte Carlo event generators, and its effect on neutron multiplicity is discussed.
Three commonly used generators are used in this paper: NEUT version 5.6.3, NuWro version 21.09.02, and GENIE version 3.04.00 with G18\_10b\_02\_11a model configuration~\cite{PhysRevD.104.072009}.
To obtain accurate excitation energy distributions to be input to NucDeEx,
the SF by Benhar {\it et al.} is adopted as the nuclear model in NEUT and NuWro.
For these two generators, the excitation energy is calculated with the energies of incoming and outgoing particles obtained from the event sample.
Since the SF has broad peaks and sometimes gives nonphysical negative excitation energies as shown in Fig.~\ref{fig:Ex}, exact energy conservation is not considered for discrete excited states, $p_{3/2}$- and $p_{1/2}$-hole states.
We need more high-precision SF considering such discrete excited states to solve this problem.
GENIE provides several sets of model configurations.
This paper uses G18\_10b\_02\_11a, which adopts the local Fermi gas and hN2018 FSI models.
GENIE also offers an effective SF model as an alternative nuclear mode.
However, it is not used because it did not show the peaks seen in Fig.~\ref{fig:Ex}.
For GENIE, the excitation energy is randomly determined according to the SF by Benhar {\it et al.} irrespective of the kinematics of the event sample.
\par
CCQE interactions of $\nu_\mu$ and $\bar{\nu}_\mu$ with monochromatic energies of 1\,GeV are investigated.
The detailed treatment of excitation energy and hole states needed to connect NucDeEx and generators are explained in Sec.~\ref{sec:ex_hole}.
Then, the nominal outputs of the generators are described in Sec.~\ref{sec:nominal}.
At last, the results of applying NucDeEx are shown in Sec.~\ref{sec:after}.

\subsection{Excitation energies and hole states} \label{sec:ex_hole}
In the case of single nucleon holes, the generators do not clarify the hole states while determining excitation energies.
Table~\ref{tab:Ex_range} shows the excitation energy range and probability of each hole state.
The hole states are determined only from the excitation energy in this application.
This simple method is not ideal because the excitation energy distribution of each hole state has a finite width and is superposed, as shown in Fig.~\ref{fig:Ex}.
A probability distribution for each hole state is needed to describe more accurately.
NEUT assumes that contributions of nucleon-nucleon correlation with excitation energies off the peak of each shell as $s_{1/2}$-hole states.
It results in a larger probability of $s_{1/2}$-hole states shown in Table~\ref{tab:Ex_range}.

\begin{table}[htbp] \centering
\caption{Probability (Prob.) and excitation energy ($E_x$) range of each hole state for $^{12}$C and $^{16}$O.
         The excitation energy ranges are set to separate peaks corresponding to each hole state of Benhar SF~\cite{PhysRevD.72.053005} shown in Fig.~\ref{fig:Ex}.
         The probability used in NEUT's original deexcitation process for $^{16}$O is also listed as a Ref.~\cite{PhysRevLett.108.052505}. 
         }
\begin{tabular*}{1.0\columnwidth}{@{\extracolsep{\fill}}lcccc} \hline \hline
    & & NEUT & \multicolumn{2}{c}{This work} \\
  Nucleus & Hole & Prob.(\%) & Prob.(\%) & $E_x$\,(MeV) \\ \hline
  $^{12}$C & $s_{1/2}$ & - & 63.9 & $E_x\geq16$ \\
           & $p_{3/2}$ & - & 36.1 & $E_x<16$ \\ \hline
  $^{16}$O & $s_{1/2}$ & 49.05 & 35.8 & $E_x\geq16$ \\
           & $p_{3/2}$ & 35.15 & 44.8 & $4\leq E_x < 16$ \\
           & $p_{1/2}$ & 15.8  & 19.4 & $E_x<4$ \\
 \hline \hline
 \label{tab:Ex_range}
\end{tabular*}
\end{table}

The excitation energy distribution in the multi-nucleon holes produced by FSI is poorly understood.
In most generators, FSI considers separation energy but not excitation energy because it is based on the Fermi gas model, which does not consider shell levels.
Hence, the generators cannot predict the excitation energy distribution in the multi-nucleon hole states.
In applying NucDeEx to the generators, the effect of FSI is neglected, only considering the excitation energy produced by the target nucleon, even in the case of multi-nucleon hole states.
It is worth mentioning that there are several studies that handle the FSI effect on excitation energy with rough approximation~\cite{HU2022137183,PhysRevD.108.112008}.
This effect is from complex many-body systems of nuclei and is quite challenging to find a reasonable description.
Exploring approaches like these studies and discussing more to reach a consensus is one of the essential issues to be addressed in the future.
Note that the excitation energy and hole state are external input parameters to NucDeEx, and any improvements in the description will require modifications on the part of the generators.

\subsection{Nominal output of generators} \label{sec:nominal}
\begin{figure*}[htbp]
\begin{minipage}[tl]{0.45\textwidth} \centering
\includegraphics[width=1.0\textwidth]{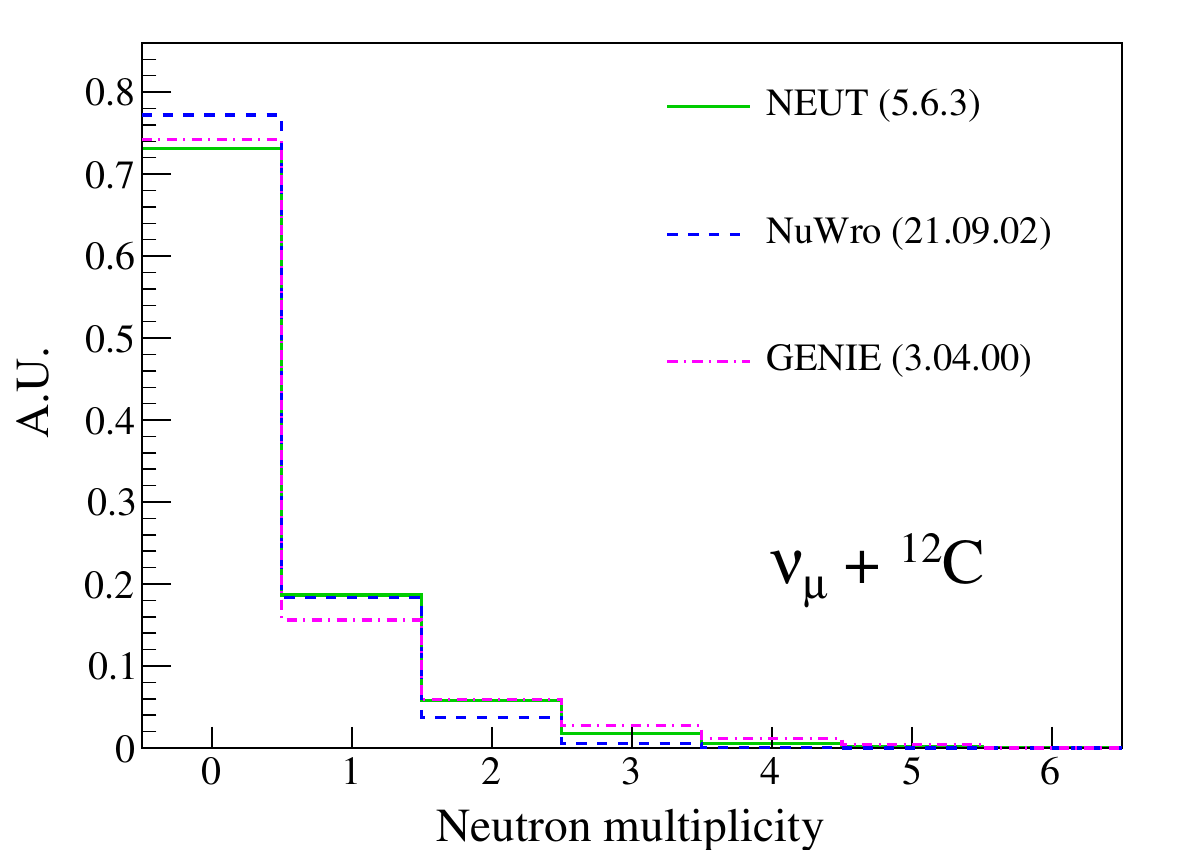}
\end{minipage}
\hspace{0.01\textwidth}
\begin{minipage}[tr]{0.45\textwidth} \centering
\includegraphics[width=1.0\textwidth]{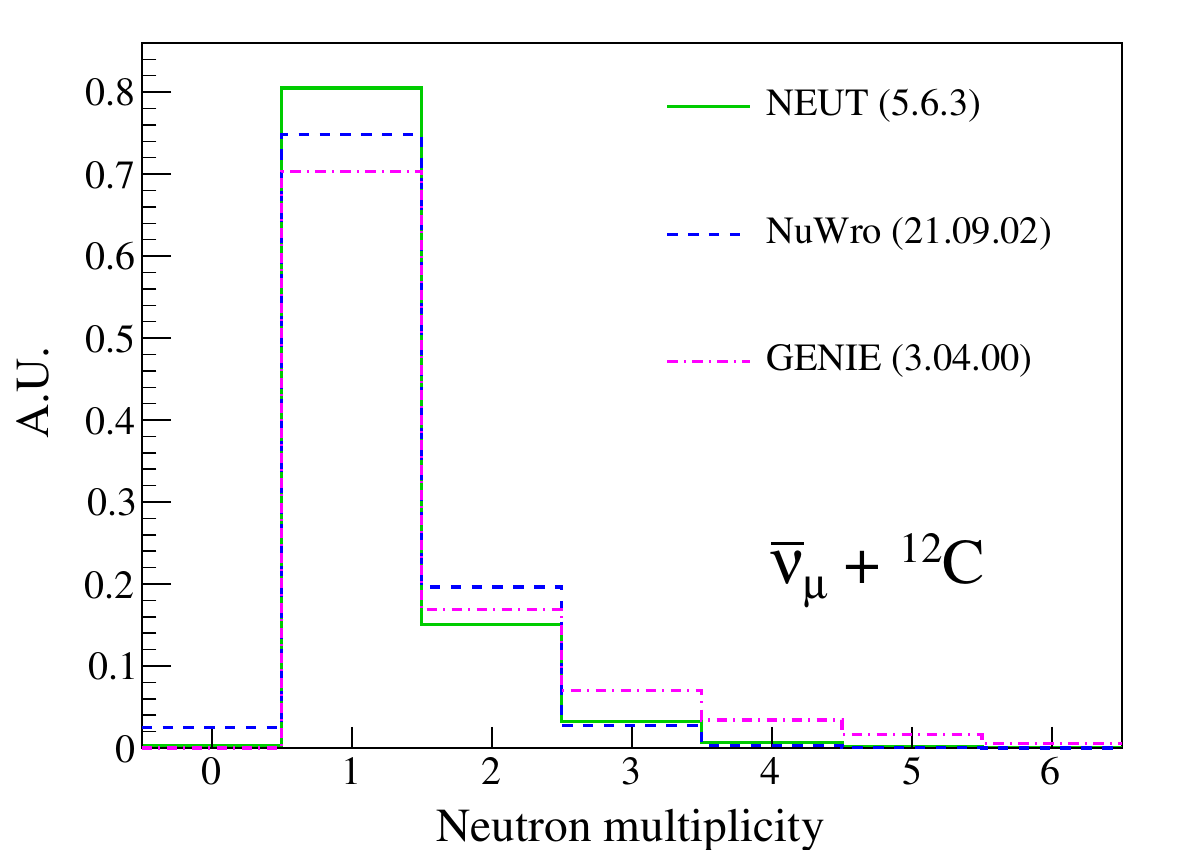}
\end{minipage}
\\
\vspace{10pt}
\begin{minipage}[bl]{0.45\textwidth} \centering
\includegraphics[width=1.0\textwidth]{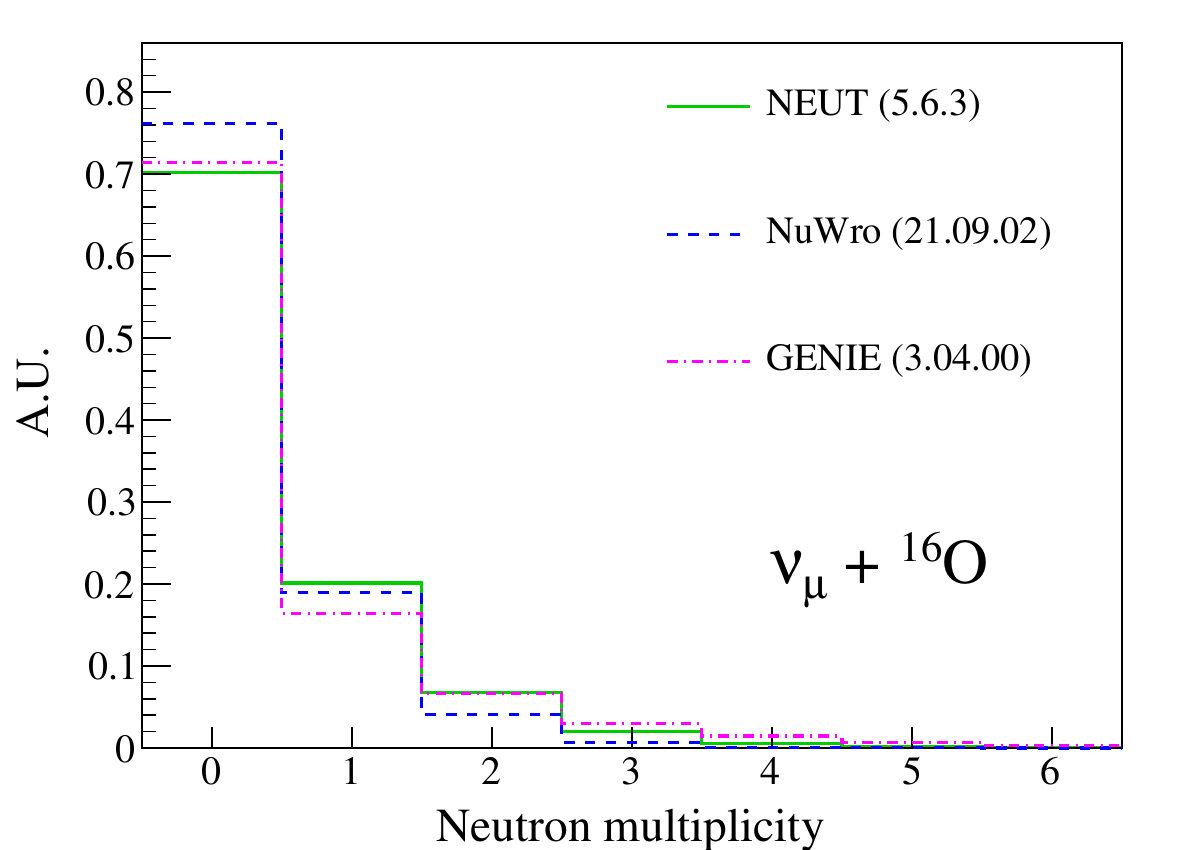}
\end{minipage}
\hspace{0.01\textwidth}
\begin{minipage}[br]{0.45\textwidth} \centering
\includegraphics[width=1.0\textwidth]{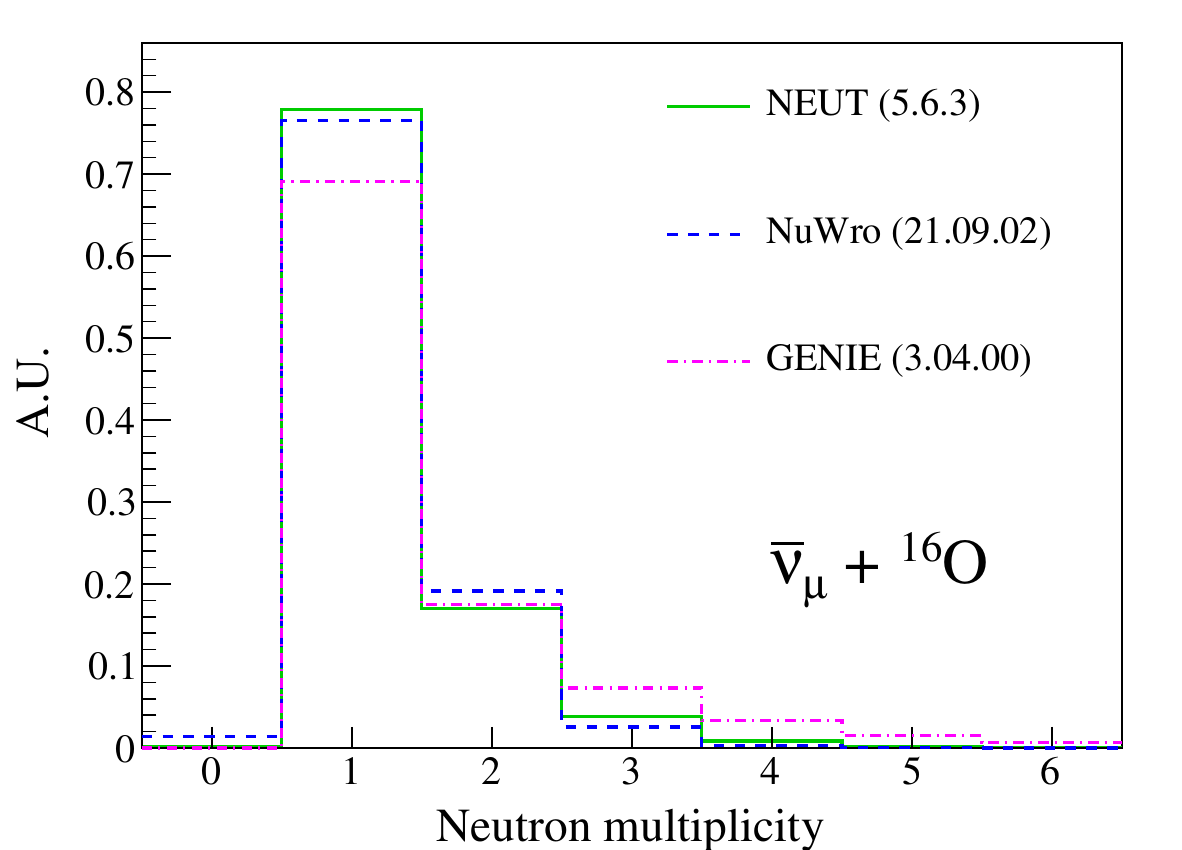}
\end{minipage}
\caption{Neutron multiplicities associated with CCQE interactions of $\nu_\mu+^{12}\text{C}$ (left top), $\bar{\nu}_\mu+^{12}\text{C}$ (right top),
         $\nu_\mu+^{16}\text{O}$ (left bottom), and $\bar{\nu}_\mu+^{16}\text{O}$ (right bottom).
         Calculations from NEUT~\cite{Hayato2021}, NuWro~\cite{PhysRevC.86.015505}, and GENIE~\cite{ANDREOPOULOS201087,PhysRevD.104.072009} are shown.
         Neutrino energy is monochromatic 1\,GeV.
         Since the deexcitation process originally implemented in NEUT is disabled,
         all calculations do not include the deexcitation process.
         }
\label{fig:nmulti_C_O}
\end{figure*}

Neutron multiplicities of the nominal output of generators are shown in Fig.~\ref{fig:nmulti_C_O}.
NEUT implements the original deexcitation process only for $^{16}$O based on Ref.~\cite{kobayashi2006deexcitation}, but it is disabled in these comparisons.
The mean neutron multiplicities are shown in Table~\ref{tab:mean_neutron_multi_12C_16O}.
These results show significant variations in neutron multiplicity of about 20\%, caused by different FSI implementations.
This means that the FSI models need to be carefully studied to predict neutron multiplicity more precisely.
NuWro (GENIE) predicts the smallest (largest) mean neutron multiplicity.
This trend is consistent with nucleon transparency discussed in Ref.~\cite{PhysRevD.104.053006}:
NuWro predicts larger transparency while GENIE gives smaller one.

\subsection{After applying NucDeEx} \label{sec:after}
\begin{figure*}[htbp]
\begin{minipage}[tr]{0.45\textwidth} \centering
\includegraphics[width=1.0\textwidth]{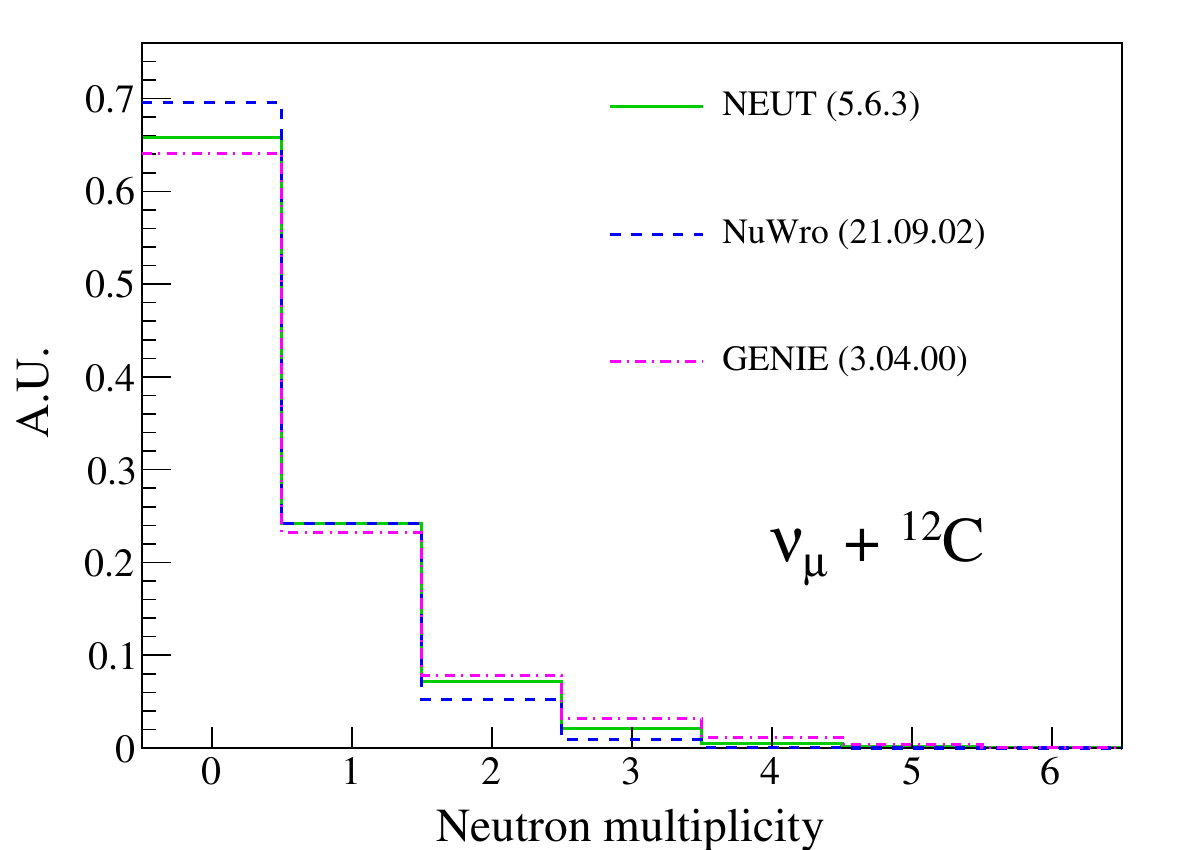}
\end{minipage}
\hspace{0.01\textwidth}
\begin{minipage}[tl]{0.45\textwidth} \centering
\includegraphics[width=1.0\textwidth]{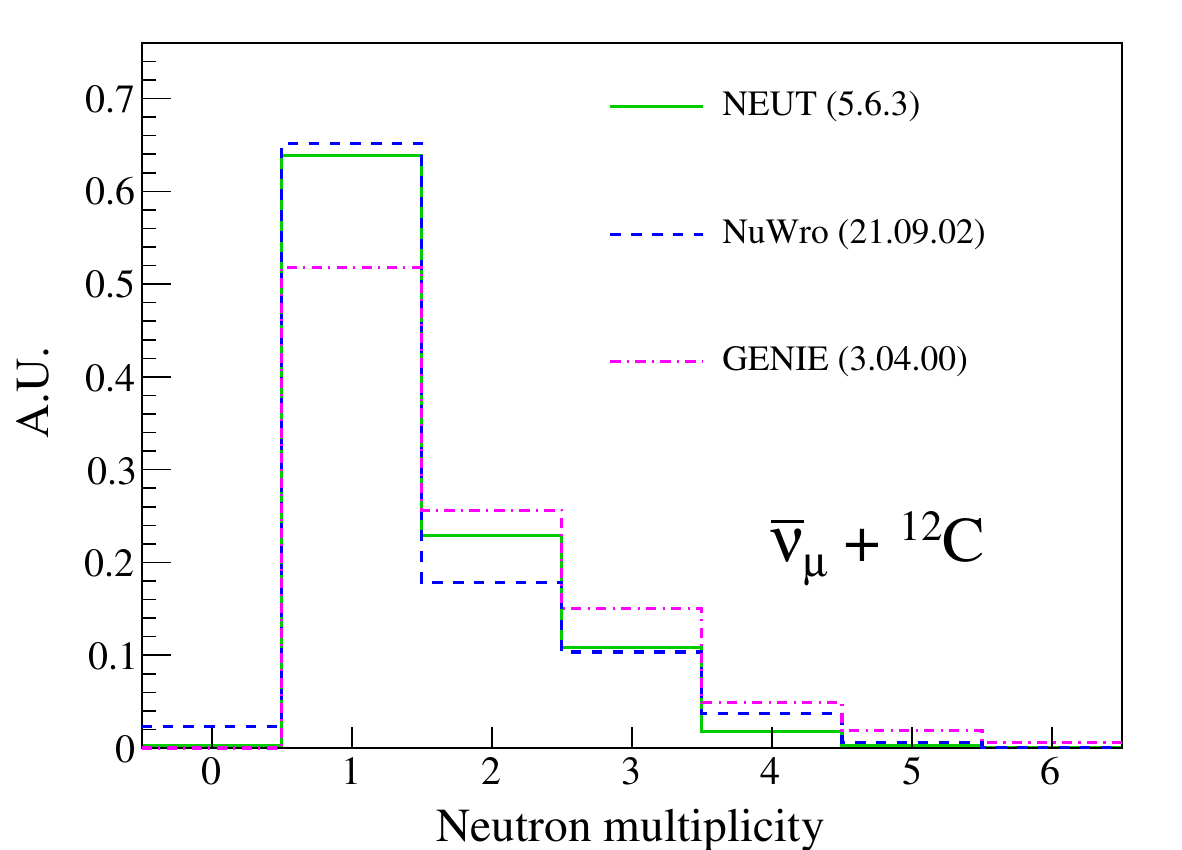}
\end{minipage}
\\
\vspace{10pt}
\begin{minipage}[br]{0.45\textwidth} \centering
\includegraphics[width=1.0\textwidth]{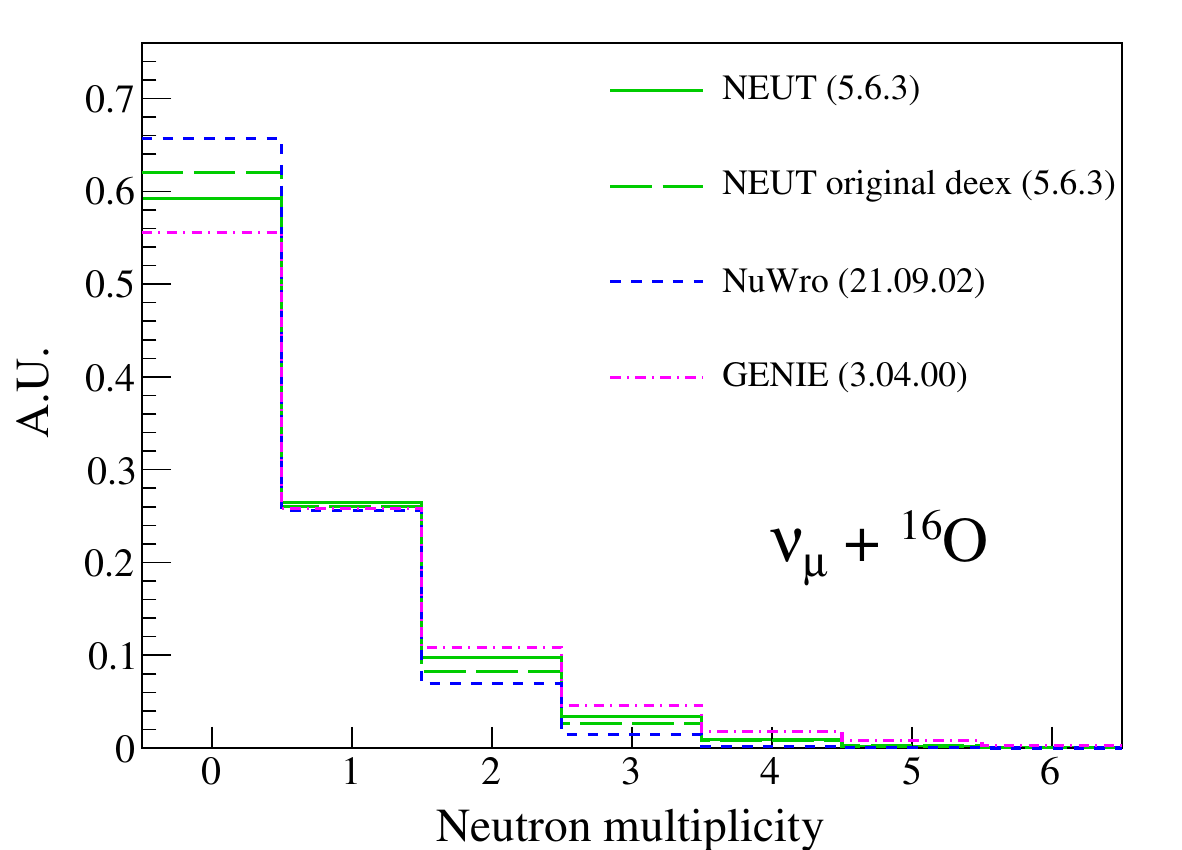}
\end{minipage}
\hspace{0.01\textwidth}
\begin{minipage}[bl]{0.45\textwidth} \centering
\includegraphics[width=1.0\textwidth]{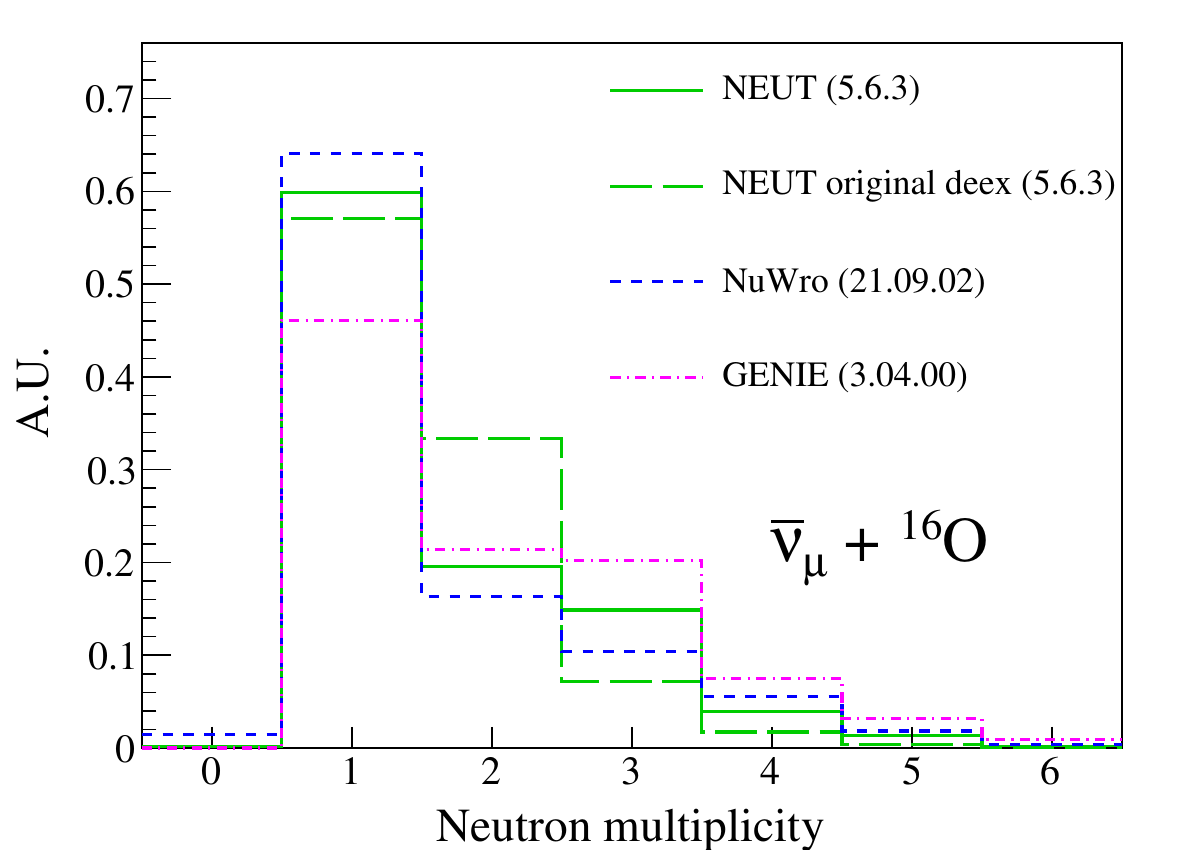}
\end{minipage}
\caption{Neutron multiplicities associated with CCQE interactions of $\nu_\mu+^{12}\text{C}$ (left top), $\bar{\nu}_\mu+^{12}\text{C}$ (right top),
         $\nu_\mu+^{16}\text{O}$ (left bottom), and $\bar{\nu}_\mu+^{16}\text{O}$ (right bottom).
         Calculations from NEUT~\cite{Hayato2021}, NuWro~\cite{PhysRevC.86.015505}, and GENIE~\cite{ANDREOPOULOS201087,PhysRevD.104.072009}
         after applying NucDeEx are shown.
         The green dashed lines represent the calculations from NEUT using its original deexcitation process, which is only applicable to $^{16}$O.
         Neutrino energy is monochromatic 1\,GeV.
         }
\label{fig:deex_nmulti_C_O}
\end{figure*}

Figure~\ref{fig:deex_nmulti_C_O} shows neutron multiplicities after applying NucDeEx.
The mean neutron multiplicities are summarized in Table~\ref{tab:mean_neutron_multi_12C_16O}.
The neutron multiplicities are increased by 20\%-30\% by deexcitation, which is equal to or greater than the differences of FSI models of generators as discussed in Sec.~\ref{sec:nominal}.
This result shows that the deexcitation processes significantly impact neutron multiplicity.
It also indicates that although the uncertainty reduction of FSI is substantial, it is even more critical to correctly treat the deexcitation process in descriptions of neutron multiplicity associated with neutrino-nucleus interactions.
\par
In the results of $^{16}$O target, calculations from NEUT using its original deexcitation process are also listed.
In the interaction of $\bar{\nu}_\mu+^{16}$O, the NEUT's calculation based on Ref.\cite{kobayashi2006deexcitation} gives a large discrepancy from the other calculations using NucDeEx.
The difference comes from the probability of each hole state shown in Table~\ref{tab:Ex_range} and the treatment of multiple neutron emissions.
The model implemented in NEUT does not consider multiple neutron emissions and overestimates single neutron emission instead.
However, multiple neutron emission is not negligible in the case of high excited states with a neutron-rich nucleus, such as $^{15}\text{N}^*$ $s_{1/2}$-hole states created by CCQE interactions of $\bar{\nu}_\mu + ^{16}\text{O}\rightarrow \mu^+ + n +^{15}\text{N}^*$.
According to NucDeEx, single (multiple) neutrons are estimated to be emitted with a probability of about 55\% (15\%) from such states.
The difference in the treatment of multiple neutron emissions results in a large difference in the probability of a single neutron emission.
It leads to a large discrepancy in the number of events in a bin with a neutron multiplicity of two in Fig.~\ref{fig:deex_nmulti_C_O}.
Since Super-Kamiokande Gadolinium has a high neutron detection efficiency of about 70\%, the data is expected to verify the difference of deexcitation models in the future.
\par
The increase in neutron multiplicity due to deexcitation depends on the generators, i.e., the FSI models, because deexcitation occurs after FSI, meaning the deexcitation depends on the residual nuclear states produced by FSI.
Even if neutrino detectors measure neutron multiplicity, separating the deexcitation and FSI contributions would be quite difficult.
The neutron travel distance of these two contributions should differ because of the dissimilar neutron energies: Neutrons emitted by deexcitation have only a few MeV.
However, the spread of the gamma ray and vertex resolution of detectors smear the information.
Therefore, it is challenging to determine which model requires modification from the neutrino experimental data only.
Our future task to be addressed is conducting non-neutrino beam experiments that can precisely measure particles around the nucleus, such as those introduced in Sec.~\ref{sec:results}, and investigating models after distinguishing between FSI and deexcitation contributions.
This task is also vital in validating the few deexcitation experimental data so far.
As well as non-neutrino beam data, creative use of neutrino data would be helpful.
The value of neutrino data can be enhanced through verification by various neutrino experiments with different detection principles or combined analysis.
Promoting both approaches, non-neutrino and neutrino experiments, is expected to lead to a more precise understanding of nuclear effects in neutrino-nucleus interactions.

\begin{table*}[htbp]
\centering
\caption{Summary of mean neutron multiplicities of CCQE interactions on $^{12}$C and $^{16}$O before and after applying NucDeEx.
         The numbers in parentheses shown in $^{16}$O denote results enabling NEUT's original deexcitation process.}
\label{tab:mean_neutron_multi_12C_16O}
\begin{tabular*}{0.75\textwidth}{@{\extracolsep{\fill}}lccccccc} \hline \hline
              & & \multicolumn{3}{c}{$\nu_\mu$} & \multicolumn{3}{c}{$\bar{\nu}_\mu$} \\
      Nucleus & & NEUT & NuWro & GENIE & NEUT & NuWro & GENIE \\ \hline
$^{12}$C & Nominal & 0.39 & 0.28 & 0.42 & 1.24 & 1.24 & 1.52 \\
& This work & 0.48 & 0.38 & 0.55 & 1.51 & 1.50 & 1.72   \\ \hline
$^{16}$O & Nominal & 0.43 (0.55) & 0.30 & 0.51 & 1.28 (1.55) & 1.24 & 1.55 \\
 & This work & 0.61 & 0.45 & 0.76 & 1.67 & 1.62 & 2.06   \\ \hline \hline
\end{tabular*}
\end{table*}

\section{Conclusion and prospects} \label{sec:conclusion}
This paper reports the construction of NucDeEx to describe the nuclear deexcitation process, which was neglected in most existing neutrino Monte Carlo event generators until recent studies.
Notable features of NucDeEx that distinguish it from similar studies are that it is open-source, can simulate deexcitation of $^{12}$C and $^{16}$O, and can be easily integrated into existing neutrino or nucleon decay Monte Carlo event generators.
The simulator comprises two components: A nuclear simulator TALYS and a kinematics simulator based on ROOT.
The quality of the simulator is evaluated by comparisons with the other theoretical predictions and hadron beam experiments.
The most critical parameters, branching ratios for neutron, agree within 20\%.
This reproducibility is adequate for an initial implementation.
The applications of NucDeEx to neutrino interactions using NEUT, NuWro, and GENIE are also demonstrated.
For the predicted neutron multiplicity, a considerable impact equal to or greater than that of FSI is pointed out.
This result shows the importance of considering the deexcitation process and the necessity of modifying the existing generators.
\par
The main future tasks for improving the accuracy are to compare NucDeEx prediction with more experiments and use them to investigate nuclear models.
As introduced in Sec.~\ref{sec:results}, only a few experiments measured the deexcitation process.
The model has room for improvement, but it is challenging to investigate how it should be modified from the limited data.
New experiments using $^{12}$C or $^{16}$O beams are worthwhile since they can detect deexcited particles with lower energy thresholds than proton beam experiments.
In addition, a detailed comparison between NucDeEx and other simulators is an interesting study to be discussed in the future.
It is worth mentioning again that a simulation study that combines NuWro and INCL FSI model coupled with a deexcitation code ABLA has been performed recently~\cite{PhysRevD.108.112008}.
INCL has also been used for hadronic interactions in Geant4 and has better reproducibility than other models~\cite{Sakai:2023Ie}.
The discussion of the deexcitation process based on the INCL FSI model is of interest.
In order to avoid dependence on a specific nuclear model, it is desirable to promote the development of deexcitation simulators for both TALYS and ABLA and to compare each other.
\par
The author provides the NucDeEx code and sample codes for applying the NucDeEx to the generators on Ref.~\cite{code}.
Note that NucDeEx supports carbon and oxygen targets but not for argon currently.
Aiming the use in liquid-argon time-projection chamber neutrino experiment, MicroBooNE~\cite{Acciarri_2017} and DUNE~\cite{acciarri2016longbaseline}, an extension to argon targets is possible in principle.
However, because of the increased number of shell levels, implementation would be more complex and difficult than with carbon or oxygen.
There is currently no concrete plan for the extension, but it might be done upon request.
The author also plans to integrate the simulator into NEUT and nucleon decay event generators used in Super-Kamiokande.
After that, NucDeEx is expected to be used in various physics studies performed in Super-Kamiokande and T2K experiment~\cite{Abe2023_T2K}.
Integrations for other generators have yet to be planned but could be easily done by the generator developers.
NucDeEx is expected to improve the prediction accuracy of neutron multiplicity associated with neutrino-nucleus interactions and nucleon decays.
The author expects that NucDeEx is capable of leading to improving various physics results using neutron multiplicity conducted in neutrino experiments in the future:
Neutrino {\it CP}-phase measurements, DSNB searches, and nucleon decay searches.

\begin{acknowledgments}
This work is supported by JSPS KAKENHI Grant No.~23KJ0319.
The author profoundly appreciates Dr.~Yoshinari Hayato for his valuable discussions.
\end{acknowledgments}

\FloatBarrier
\bibliography{bib}

\end{document}